\documentclass[main]{aa}
\usepackage{graphicx}
\usepackage{txfonts}

\def\Lsun{L$_\odot$}
\def\Msun{M$_\odot$}

\def\Cii{[C\,{\sc ii}]}
\def\Ci{[C\,{\sc i}]}

\def\kms{km\,s$^{-1}$}

\def\Kkmspc{K~km\,s$^{-1}$\,pc$^2$}
\def\lsim{\mathrel{\rlap{\lower 3pt \hbox{$\sim$}} \raise 2.0pt \hbox{$<$}}}
\def\gsim{\mathrel{\rlap{\lower 3pt \hbox{$\sim$}} \raise 2.0pt \hbox{$>$}}}

\begin{document}

\authorrunning{Decarli et al.}
\titlerunning{Molecular gas in $z\sim 6$ quasars}

\title{Molecular gas in $z\sim 6$ quasar host galaxies}
\author{
Roberto Decarli\inst{1},
Antonio Pensabene\inst{1,2,3},
Bram Venemans\inst{4,5},
Fabian Walter\inst{4,6},
Eduardo Ba\~{n}ados\inst{4},
Frank Bertoldi\inst{7},
Chris L.~Carilli\inst{6},
Pierre Cox\inst{8,9},
Xiaohui Fan\inst{10},
Emanuele Paolo Farina\inst{11},
Carl Ferkinhoff\inst{12},
Brent A.~Groves\inst{13,14},
Jianan Li\inst{15,16,17},
Chiara Mazzucchelli\inst{18},
Roberto Neri\inst{19},
Dominik A.~Riechers\inst{20},
Bade Uzgil\inst{21},
Feige Wang\inst{10},
Ran Wang\inst{15},
Axel Weiss\inst{22},
Jan Martin Winters\inst{19},
Jinyi Yang\inst{10}
}
\institute{
INAF -- Osservatorio di Astrofisica e Scienza dello Spazio di Bologna, via Gobetti 93/3, I-40129, Bologna, Italy. \email{ roberto.decarli@inaf.it} \and
Dipartimento di Fisica e Astronomia, Alma Mater Studiorum, Universit\'{a} di Bologna, Via Gobetti 93/2, I-40129 Bologna, Italy. \and
Dipartimento di Fisica ``G. Occhialini'', Universit\`a degli Studi di Milano-Bicocca, Piazza della Scienza 3, I-20126, Milano, Italy. \and
Max-Planck Institut f\"{u}r Astronomie, K\"{o}nigstuhl 17, D-69117, Heidelberg, Germany. \and
Leiden Observatory, Leiden University, P.O. Box 9513, NL-2300 RA Leiden, The Netherlands. \and
National Radio Astronomy Observatory, Pete V.\,Domenici Array Science Center, P.O.\, Box O, Socorro, NM, 87801, USA.\and
Argelander-Institute for Astronomy, University of Bonn, Auf dem H\"{u}gel 71, D-53121 Bonn, Germany. \and
Sorbonne Universit{\'e}, UPMC Universit{\'e} Paris 6 \& CNRS, UMR 7095.\and
Institut d'Astrophysique de Paris, 98b boulevard Arago, 75014 Paris, France. \and
Steward Observatory, University of Arizona, 933 N. Cherry St., Tucson, AZ  85721, USA. \and
Gemini Observatory, NSF's NOIRLab, 670 N A'ohoku Place, Hilo, Hawai'i 96720, USA. \and
Department of Physics, Winona State University, Winona, MN 55987, USA. \and
ICRAR M468, The University of Western Australia, 35 Stirling Hwy, Crawley, WA 6009, Australia.\and
Research School of Astronomy and Astrophysics (RSAA), Australian National University, ACT 2611, Australia. \and
Department of Astronomy, School of Physics, Peking University, 5 Yiheyuan Road, Haidian District, Beijing, 10087, China. \and
Kavli Institute of Astronomy and Astrophysics at Peking University, 5 Yiheyuan Road, Haidian District, Beijing 100871, China. \and
Department of Astronomy, Tsinghua University, Beijing 100084, China. \and	
European Southern Observatory, Alonso de C\'{o}rdova 3107, Vitacura, Regi\'{o}n Metropolitana, Chile. \and
Institut de Radioastronomie Millim\'{e}trique (IRAM), 300 Rue de la Piscine, 38400 Saint-Martin-d'H\`{e}res, France. \and
Department of Astronomy, Cornell University, Space Sciences Building, Ithaca, NY 14853, USA. \and
Astronomy Department, California Institute of Technology, MC249-17, Pasadena, California 91125, USA. \and
Max-Planck-Institut f\"{u}r Radioastronomie, Auf dem H\"{u}gel 69, D-53121 Bonn, Germany.
}

\date{September 2021}
\abstract{We investigate the molecular gas content of $z\sim 6$ quasar host galaxies using the Institut de Radioastronomie Millim\'{e}trique / Northern Extended Millimeter Array. We target the 3\,mm dust continuum, and the line emission from CO(6--5), CO(7--6), \Ci{}$_{2-1}$ in 10 infra-red--luminous quasars that have been previously studied in their 1\,mm dust continuum and \Cii{} line emission. We detect CO(7--6) at various degrees of significance in all the targeted sources, thus doubling the number of such detections in $z\sim 6$ quasars. The 3\,mm to 1\,mm flux density ratios are consistent with a modified black body spectrum with a dust temperature $T_{\rm dust}\sim 47$\,K and an optical depth $\tau_{\nu}=0.2$ at the \Cii{} frequency. Our study provides us with four independent ways to estimate the molecular gas mass, $M_{\rm H2}$, in the targeted quasars. This allows us to set constraints on various parameters used in the derivation of molecular gas mass estimates, such as the mass per luminosity ratios $\alpha_{\rm CO}$ and $\alpha_{\rm [CII]}$, the gas-to-dust mass ratio $\delta_{\rm g/d}$, and the carbon abundance [C]/H$_2$. Leveraging either on the dust, CO, \Ci{}, or \Cii{} emission yields mass estimates of the entire sample in the range $M_{\rm H2}\sim 10^{10}$ to $10^{11}$\,\Msun{}. We compare the observed luminosities of dust, \Cii{}, \Ci{}, and CO(7--6) with predictions from photo-dissociation and X-ray dominated regions. We find that the former provide  better model fits to our data, assuming that the bulk of the emission arises from dense ($n_{\rm H}>10^4$\,cm$^{-3}$) clouds with a column density $N_{\rm H}\sim10^{23}$\,cm$^{-2}$, exposed to a radiation field with intensity $G_0\sim 10^3$ (in Habing units). Our analysis reiterates the presence of massive reservoirs of molecular gas fueling star formation and nuclear accretion in $z\sim 6$ quasar host galaxies. It also highlights the power of combined 3\,mm and 1\,mm observations for quantitative studies of the dense gas content in massive galaxies at cosmic dawn.
}

\keywords{galaxies: high-redshift --- galaxies: evolution --- galaxies: ISM --- 
galaxies: star formation ---  quasars: emission lines}
\maketitle

\section{Introduction}

Quasars residing in the so-called `cosmic dawn' ($z\gsim6$, age of the Universe $<$1\,Gyr) were first discovered at the turn of the century \citep[e.g.,][]{fan00,fan03}. To date, we know $\sim 400$ quasars at $z>5.5$, including three at $z>7.5$ \citep{banados18,yang20,wang21}. Multi-wavelength campaigns, in particular at sub-millimeter and millimeter (hereafter, sub-mm and mm) wavelengths, revealed that these early quasars reside in extremely active galaxies, experiencing intense starbursts \citep[with star formation rates SFR=100-1000\,\Msun{}\,yr$^{-1}$;][]{bertoldi03,wang08a,wang08b,drouart14,leipski14,venemans18} and fast, radiatively-efficient black hole accretion \citep[$\dot{M}_{\rm BH}>10$\,\Msun{}\,yr$^{-1}$;][]{derosa14,schindler20,yang21}. This rapid gas consumption depletes the immense reservoirs of molecular gas \citep[$M_{\rm H2}>10^{10}$\,\Msun{};][]{walter03,wang10,venemans17a,decarli18} of the host galaxies, that can be refueled via accretion of cool gas from the circumgalactic medium \citep[e.g.,][]{farina19,drake19} and/or via mergers with gas--rich galaxies \citep[e.g.,][]{trakhtenbrot17,decarli17,decarli19,vito19}.

\begin{table*}
\caption{The sample of quasars studied in this work. (1) Full quasar name. (2) Abbreviated target name. (3--4) Target coordinates. (5) Redshift. (6) Continuum flux density at the observed frequency of the \Cii{} line. (7) Total infrared luminosity, $L_{\rm IR}$, integrated between 8--1000 $\mu$m (rest frame), computed as described in Sec.~\ref{sec_fits}. (8) \Cii{} luminosities. (9) References for \Cii{} and IR measurements.
}\label{tab_sample}
\vspace{-8mm}
\begin{center}
\begin{tabular}{ccccccccc}
\hline
Full name              & Target       &  R.A.        & Dec.         & $z$    &  $F_{\rm cont}$ (1\,mm) &$L_{\rm IR}$        & $L_{\rm [CII]}$   & Ref  \\
                       &              & (J2000.0)    & (J2000.0)    &        &  [mJy]		       &[$10^{12}$\,\Lsun{}] & [$10^9$\,\Lsun{}] &      \\
 (1)                   & (2)          & (3)          & (4)          & (5)    & (6)                     & (7)                 & (8)               & (9)  \\
\hline
 PSO~J036.5078+03.0498 & PJ036+03     & 02:26:01.876 & +03:02:59.39 & 6.5412 & $2.5  \pm0.5  $ & $6.5$  	& $5.55_{-0.64}^{+0.64}$  & 3	\\ 
       SDSS~J0338+0021 & J0338+0021   & 03:38:29.310 & +00:21:56.30 & 5.0267 & $2.98 \pm0.05 $ & $12.0^\dagger$	& $5.69_{-0.83}^{+0.83}$  & 2	\\ 
PSO~J159.2257--02.5438 & PJ159--02    & 10:36:54.191 &--02:32:37.94 & 6.3809 & $0.65 \pm0.03 $ & $1.6$  	& $1.19_{-0.07}^{+0.07}$  & 6	\\ 
       VIK~J1048--0109 & J1048--0109  & 10:48:19.086 &--01:09:40.29 & 6.6759 & $2.84 \pm0.03 $ & $7.7$  	& $2.77_{-0.07}^{+0.08}$  & 6	\\ 
       DELS~J1104+2134 & J1104+2134   & 11:04:21.580 & +21:34:28.85 & 6.7662 & $1.8  \pm0.05 $ & $5.0$  	& $1.82_{-0.28}^{+0.28}$  & 7	\\ 
       ULAS~J1319+0950 & J1319+0950   & 13:19:11.302 & +09:50:51.49 & 6.1331 & $5.23 \pm0.10 $ & $10.7^\dagger$	& $4.22_{-0.58}^{+0.58}$  & 1,2 \\ 
      SDSS~J2054--0005 & J2054--0005  & 20:54:06.481 &--00:05:14.80 & 6.0391 & $2.98 \pm0.05 $ & $8.0^\dagger$	& $1.89_{-0.11}^{+0.11}$  & 1,2 \\ 
       VIMOS2911001793 & J2219+0102   & 22:19:17.217 & +01:02:48.90 & 6.1492 & $0.766\pm0.047$ & $1.8$  	& $2.48_{-0.13}^{+0.13}$  & 5	\\ 
PSO~J338.2298+29.5089  & PJ338+29     & 22:32:55.150 & +29:30:32.23 & 6.6660 & $0.972\pm0.215$ & $2.6$		& $1.89_{-0.92}^{+1.00}$  & 4	\\ 
PSO~J359.1352--06.3831 & PJ359--06    & 23:56:32.455 &--06:22:59.26 & 6.1722 & $0.87 \pm0.08 $ & $2.1$  	& $2.42_{-0.16}^{+0.13}$  & 6	\\ 
\hline
\end{tabular}
\end{center}
\vspace{-5mm}
\tablebib{
1-\citet{wang13}, 
2-\citet{leipski14},
3-\citet{banados15}, 
4-\citet{mazzucchelli17}, 
5-\citet{willott17}, 
6-\citet{decarli18}, 
7-F.\ Wang et al. (in prep).\\
$^\dagger$ IR luminosity based on SED fits from the literature.}
\end{table*}

\begin{table*}
\caption{Summary of the observations presented in this paper. (1) Target name. (2) ID of the observing program. (3) Array configuration. (4) Number of visibilities in the final cubes. For sources with independent datasets for the lower and upper side bands, both values are quoted. (5--6) Synthesized beams in the lower and upper side bands, in arcsec. (7--8) Continuum rms in the two side bands. (9--10) Median channel rms, per 50\,\kms{} channel, in the two side bands.}\label{tab_obs}
\vspace{-4mm}
\begin{center}
\begin{tabular}{cccccccccc}
\hline
Target         & Program ID   & Config.      & \# of visibilities & \multicolumn{2}{c}{Beam}   & \multicolumn{2}{c}{$\sigma_{\rm cont}$}    & \multicolumn{2}{c}{$\sigma_{\rm chn}$} \\
               &              &              &              & \multicolumn{2}{c}{[$''\times ''$]} & \multicolumn{2}{c}{[$\mu$Jy\,beam$^{-1}$]} & \multicolumn{2}{c}{[mJy\,beam$^{-1}$]} \\
               &              &              &              & LSB      & USB             & LSB           & USB                        & LSB                 & USB              \\
 (1)           & (2)          & (3)          & (4)          & (5)      & (6)             & (7)           & (8)                        & (9)                 & (10)             \\
\hline
 PJ036+03      & S15DA, S17CD & 5D--8D       & 22317,22440  & $5.7\times3.6$ & $6.1\times3.5$ & 23	    & 38	      & 0.30		    & 0.37	       \\
 J0338+0021    & S052, X04D   & 5D           & 7200         &                & $4.7\times3.3$ &		    & 200 	      &      		    & 0.88 	       \\  
 PJ159--02     & S19DM        & 10C          & 7170         & $3.8\times1.4$ & $3.2\times1.2$ & 44	    & 48   	      & 0.29     	    & 0.44	       \\  
 J1048--0109   & S19DM        & 8C,10C,9D    & 10170        & $4.3\times1.8$ & $3.6\times1.6$ & 40          & 52	      & 0.32  		    & 0.39	       \\
 J1104+2134    & S19DM        & 9D           & 12660        & $5.5\times3.8$ & $4.8\times3.1$ & 18	    & 19	      & 0.27  		    & 0.31	       \\
 J1319+0950    & S19DM        & 9D           & 8310         & $5.7\times3.2$ & $5.0\times2.7$ & 18	    & 28	      & 0.23 		    & 0.41	       \\
 J2054--0005   & X04D         & 6C,5D--6D    & 16881        &                & $4.0\times2.6$ &      	    & 65   	      &      		    & 0.73	       \\
 J2219+0102    & S18DM, S19DM & 8D-9D        & 22020,8640   & $6.5\times3.2$ & $5.4\times2.9$ & 20          & 22	      & 0.22  		    & 0.38            \\
 PJ338+29      & S18DM        & 8D           & 8670         & $4.6\times3.5$ & $3.9\times3.1$ & 25	    & 31   	      & 0.37  		    & 0.45	       \\
 PJ359--06     & S19DM        & 9D           & 9961         & $7.3\times3.5$ & $6.4\times2.9$ & 23	    & 25   	      & 0.32  		    & 0.48	       \\
\hline
\end{tabular}
\end{center}
\end{table*}

Both star formation and black hole accretion release huge amounts of energy which could affect the evolution of the host galaxies: Mechanical feedback (via winds or jets) can displace and even remove the gas reservoir; radiative feedback can heat-up and excite the interstellar medium (ISM) thus slowing down or preventing cooling and gas fragmentation \citep[see][for reviews on this topic]{fabian12,heckman14,king15,somerville15}. A quantitative description of the cold ($T<100$\,K) ISM, poised to constrain the mass, density, temperature, and excitation mechanism of the molecular gas in the quasar hosts, would shed light on the interplay between the ISM, star formation and nuclear activity. 

\begin{figure*}
\begin{center}
\includegraphics[width=0.49\textwidth]{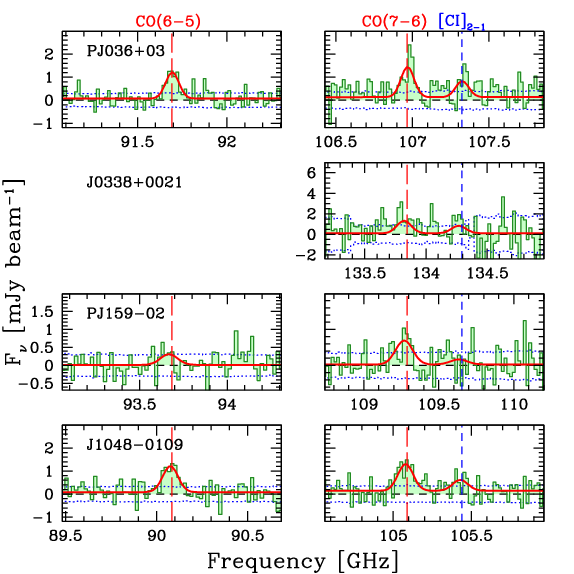}
\includegraphics[width=0.49\textwidth]{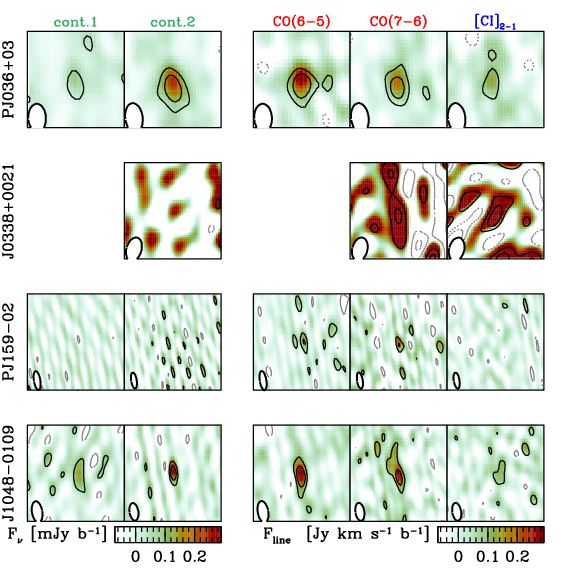}\\
\end{center}
\caption{IRAM PdBI and NOEMA spectra and images of CO(6--5), CO(7--6), and \Ci{} in quasar host galaxies at $z\sim6$. In the spectra, the dotted histograms mark the +1/-1 $\sigma$ range in the rms noise. The vertical red long-dashed / blue short-dashed lines mark the expected frequencies of the CO / \Ci{} transitions. The thick red lines show the best fit model of the spectra. In the right panels, the two continuum images are obtained by integrating over all of the line--free channels in the lower (1) and upper (2) side bands. The line maps are obtained by integrating over the observed or expected line width (see text for details). All the panels are $20''\times20''$ wide centered on the quasars. North is on the top, East to the left-hand side. Black/grey contours show the $\pm2$,4,8,16-$\sigma$ isophotes. The synthesized beam is shown as a white ellipse at the bottom--left corner of the images. }\label{fig_spc1}
\end{figure*}


\begin{figure*}
\begin{center}
\includegraphics[width=0.49\textwidth]{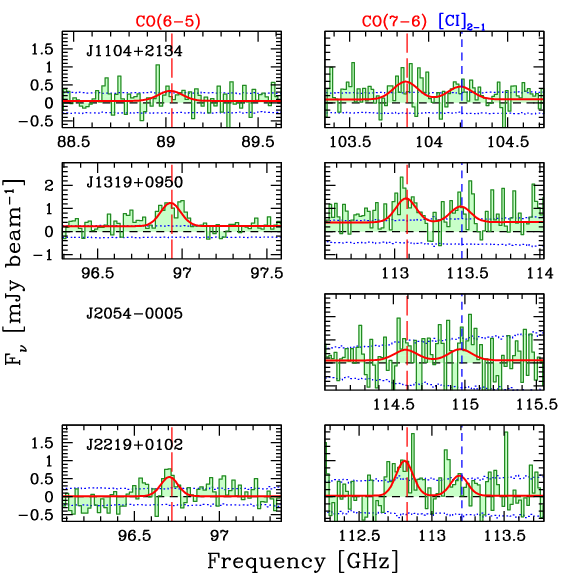}
\includegraphics[width=0.49\textwidth]{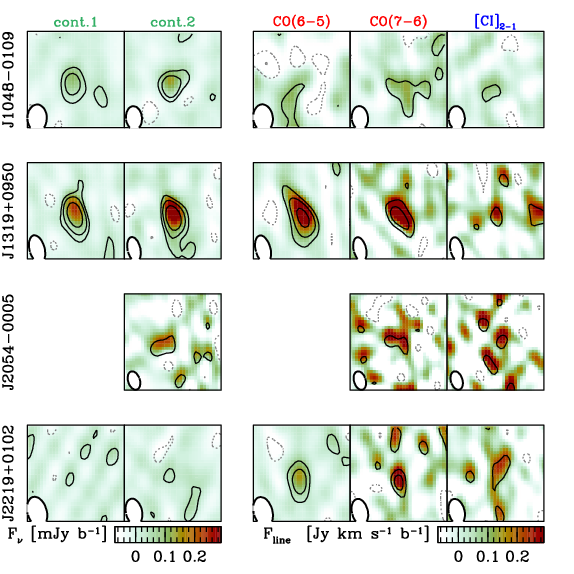}\\
\includegraphics[width=0.49\textwidth]{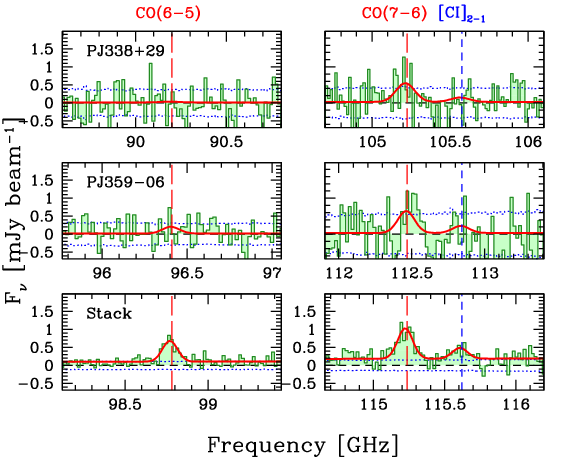}
\includegraphics[width=0.49\textwidth]{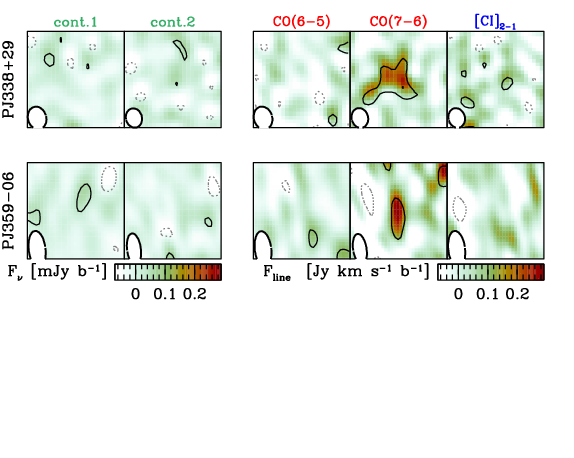}\\
\end{center}
\caption{Continued from Fig.~\ref{fig_spc1}. In the last row, we show the stacked spectra of CO(6--5), CO(7--6) and \Ci{} obtained as described in Sec.~\ref{sec_fits}.}\label{fig_spc3}
\end{figure*}

In distant galaxies, the molecular gas mass, $M_{\rm H2}$, is usually inferred based on photometry covering the Rayleigh-Jeans part of the dust continuum emission \citep[see, e.g.,][]{groves15,scoville17}, or via far-infrared emission lines of species such as the carbon monoxide, $^{12}$C$^{16}$O (hereafter, CO; see, e.g., \citealt{walter03}, \citealt{wang11}, as well as the review in \citealt{carilli13}), neutral carbon \citep[in particular the fine-structure transitions \Ci{}$_{1-0}$ and \Ci{}$_{2-1}$; see, e.g.,][]{popping17,valentino18,boogaard20}, or singly-ionized carbon \citep[via the bright \Cii{}$_{3/2-1/2}$ line; see][]{venemans17b,zanella18}. All of these methods rely on a set of assumptions and conversion factors: e.g., the gas-to-dust ratio, $\delta_{\rm g/d}$ \citep[see, e.g.,][]{berta16} or the $\alpha_{\rm CO}$ factor that links the luminosity of the CO(1-0) ground transition with the associated molecular gas mass \citep[see][for a review]{bolatto13}. These factors carry implicit dependencies on the metallicity and other properties of the galaxy. Comparisons of $M_{\rm H2}$ estimates obtained via different prescriptions are pivotal to pin down systematics in the mass determinations. For instance, \citet{dunne21} used a combination of dust continuum, CO(1--0) and \Ci{}$_{1-0}$ line emission to gauge the molecular gas mass in a sample of massive galaxies at $z=0.35$. Because four unknown quantities (molecular gas mass, carbon abundance, $\alpha_{\rm CO}$, $\delta_{\rm g/d}$) are constrained with three observables, only relative values can be directly measured; however, by using additional information (e.g., from their sample variance), mean values for the calibration factors can be inferred. Similarly, \citet{sommovigo21,sommovigo22} proposed to use a combination of nested scaling relations to constrain the dust temperature, $T_{\rm d}$, from the \Cii{} line luminosity.

The combination of different diagnostics of the cold ISM also reveal some precious information on the physical conditions of the gas. For instance, models of Photo-Dissociation Regions (PDRs) or X-ray Dominated Regions (XDRs) account for various chemical and radiative processes that take place in a gas cloud that is illuminated by an external radiation field in order to predict the emerging intensity of various lines. The former set of models assume that the excitation is driven by the UV radiation from young stars, and they are thus best suited to describe the emission of molecular clouds invested by the radiative output of an on-going starburst. The latter models assume the presence of a strong X-ray radiation field as the driver of the cold gas heating and excitation. These models are hence best suited to account for the quasar contribution to the excitation budget, although both PDR and XDR models have been extensively used in a wide range of astrophysical applications \citep[e.g.,][]{tielens85,maloney96,meijerink07,goicoechea19,vallini19}. By exploiting differences in the predicted line emission in PDRs versus XDRs, we can infer hints on the role of different radiative mechanisms (star formation or nuclear activity) that may drive the excitation of the cold ISM. 
Dust, \Cii{}, \Ci{}, and CO(7--6) form an observationally convenient suite of diagnostics, as with only two frequency settings (one covering the \Cii{}, and one covering \Ci{} and CO(7--6), together with their respective underlying continuum) we can secure a rich collection of tracers of the cold interstellar medium. 

In this study, we present the first systematic campaign targeting CO(7--6) and \Ci{} in a sample of ten $z\sim 6$ quasars. In addition, the unprecedented bandwidth capabilities of new mm receivers together with upgraded correlators enable to observe CO(6--5) as well in the same frequency setting as CO(7--6) and \Ci{}, for certain redshift windows. We used the Plateau de Bure Interferometer (PdBI), later upgraded to the NOrthern Extended Millimeter Array (NOEMA) of the Institut de Radioastronomie Millim\'{e}trique (IRAM) to secure CO(6--5), CO(7--6) and \Ci{} coverage in the 3mm atmospheric window. All of these quasars have been previously detected in the \Cii{} and dust continuum emission. The paper is structured as follows: in sec.~\ref{sec_obs}, we introduce the sample, the observations, and the data reduction. In sec.~\ref{sec_results}, we present our results, our estimates of $M_{\rm H2}$ based on various methods, and our constraints on the physical properties of the ISM. We draw our conclusions in sec.~\ref{sec_conclusions}. Throughout this paper we assume a $\Lambda$CDM cosmology, with $H_0=70$ \kms{}\,Mpc$^{-1}$, $\Omega_{\rm m}=0.3$, and $\Omega_{\Lambda}=0.7$, and a Kroupa initial mass function to compute star formation rates (SFR)\footnote{For reference, SFR estimates based on a \citet{chabrier03} initial mass function are practically unchanged, while adopting a \citet{salpeter55} initial mass function would lead to $\sim 15$\% higher SFR estimates \citep{kennicutt12}.}. In this cosmological framework, the scale distance at $z$=$6.00$ is 5.713 kpc\,arcsec$^{-1}$, the luminosity distance is $D_{\rm L}$=57.742\,Gpc, and the age of the Universe is 914 Myr. The notation $\nu$ refers to rest--frame frequencies.

\section{Observations and data processing}\label{sec_obs}

\subsection{Sample}\label{sec_sample}

Our goal is to capitalize on the combination of \Cii{}, \Ci{}, CO, and the underlying dust continuum emission to study the molecular medium in quasar host galaxies at the dawn of cosmic time. To this purpose, we selected all of the quasars with previously published \Cii{} detections. At the time of designing the main part of this program (March 2019), out of $\sim 150$ quasars known at $z\gsim 6$, fourty-six were detected in \Cii{}. From these, only nine had previously been observed in their CO(7--6) emission: one  (PSO~J183.1124+05.0926) is currently unpublished (R.~Decarli et al.\ in prep.), and two show no line detection (ULAS~J1120+0641, ULAS~J1342+0928; see \citealt{venemans17b} and \citealt{novak19}). The remaining six quasars (SDSS~J1148+5251, VIK~J0109--3047, VIK~J0305--3150, VIK~J2348--3054, SDSS~J0100+2802, UHS~J0439+1634) have CO(7--6) observations published in \citet{riechers09}, \citet{venemans17a}, \citet{wang19}, and \citet{yang19}. We removed all of the sources previously observed in CO(7--6) from our selection. We also removed 12 additional quasars that reside at $5.9\lsim z\lsim 6.1$, as at these redshifts the CO(7--6) + \Ci{} lines are significantly affected by poorer atmospheric transmission. We also removed six quasars that have IR luminosity $L_{\rm IR}<10^{12}$\,\Lsun{}, i.e., that are likely too faint at IR wavelengths for our survey. Finally, we drop 11 quasars that lie at declination $<-10^\circ$ for the sake of observability from NOEMA. In addition to this sample, we include previously unpublished observations of the quasars SDSS~J0338+0021 ($z$=$5.0267$) and SDSS~J2054-0005 ($z$=$6.0391$). The resulting list of targets thus consists of ten quasars (see Table~\ref{tab_sample}).

\begin{table*}
\caption{Results from the spectral fits. (1) Target name. The last entry reports the results on the stacked spectrum (see Sec.~\ref{sec_fits}). (2) Fitted redshift, based on the CO and \Ci{} lines presented here. (3) Continuum flux density at the observed frequency of the CO(7--6) line. (4) Fitted line full width at half maximum (for all the fitted CO and \Ci{} lines in this study). (5--7) Fitted line integrated fluxes. Limits are at 3-$\sigma$ significance. (8--10) Inferred line luminosities.}\label{tab_fits}
\vspace{-4mm}
\begin{center}
\begin{tabular}{cccccccccc}
\hline
Target      &$z_{\rm CO, [CI]}$& $F_{\rm cont}$ (3\,mm)	&  FWHM	   & \multicolumn{3}{c}{$F_{\rm line}$ [Jy\,km\,s$^{-1}$]} & \multicolumn{3}{c}{$L'$ [$10^9$ \Kkmspc]}    \\
            &                  & [mJy]			&  [\kms]	   & CO(6--5)			   & CO(7--6)		   & [C{\sc i}] 	       & CO(6--5)	      & CO(7--6)	   & [C{\sc i}] 	\\
 (1)        & (2)	       &  (3)                   & (4)		    & (5)			    & (6)		    & (7)                 & (8)                 & (9)                & (10) \\
\hline
PJ036+03    &$6.5410\pm0.0003$ & $0.131\pm0.020$ & $ 283_{-21}^{+24}$ &  $0.345_{-0.031}^{+0.029}$ & $0.397_{-0.045}^{+0.039}$ & $0.211_{-0.042}^{+0.048}$   & $12.7_{-1.2}^{+1.1}$ & $10.7_{-1.2}^{+1.1}$ & $5.7_{-1.1}^{+1.3}$  \\
J0338+0021  &$5.0278\pm0.0003$ & $<0.600       $ & $ 309_{-37}^{+43}$ &  ---			   & $0.399_{-0.057}^{+0.065}$ & $0.241_{-0.063}^{+0.060}$   &  ---		    &  $7.3_{-1.0}^{+1.2}$ & $4.4_{-1.2}^{+1.1}$  \\
PJ159--02   &$6.3822\pm0.0004$ & $<0.097       $ & $ 366_{-51}^{+56}$ &  $0.100_{-0.034}^{+0.030}$ & $0.262_{-0.048}^{+0.051}$ & $<0.150$		     &  $3.5_{-1.2}^{+1.1}$ &  $6.8_{-1.2}^{+1.3}$ & $<3.9$		  \\
J1048--0109 &$6.6766\pm0.0003$ & $0.146\pm0.032$ & $ 327_{-30}^{+32}$ &  $0.340_{-0.037}^{+0.038}$ & $0.403_{-0.039}^{+0.037}$ & $0.164_{-0.051}^{+0.046}$   & $12.9_{-1.4}^{+1.4}$ & $11.2_{-1.1}^{+1.0}$ & $4.5_{-1.4}^{+1.3}$  \\
J1104+2134  &$6.7672\pm0.0004$ & $0.096\pm0.013$ & $ 513_{-67}^{+58}$ &  $0.131_{-0.037}^{+0.031}$ & $0.270_{-0.068}^{+0.043}$ & $0.194_{-0.048}^{+0.045}$   &  $5.1_{-1.4}^{+1.2}$ &  $7.7_{-1.9}^{+1.2}$ & $5.5_{-1.3}^{+1.3}$  \\
J1319+0950  &$6.1336\pm0.0003$ & $0.403\pm0.015$ & $ 428_{-32}^{+35}$ &  $0.397_{-0.038}^{+0.028}$ & $0.474_{-0.058}^{+0.060}$ & $0.313_{-0.044}^{+0.048}$   & $13.3_{-1.3}^{+0.9}$ & $11.7_{-1.4}^{+1.5}$ & $7.6_{-1.1}^{+1.2}$  \\
J2054--0005 &$6.0397\pm0.0003$ & $<0.194       $ & $487_{-115}^{+108}$&  ---			   & $0.235_{-0.075}^{+0.093}$ & $<0.261$		     &  ---		    &  $5.6_{-1.8}^{+2.2}$ & $<6.3$		  \\
J2219+0102  &$6.1503\pm0.0004$ & $<0.045       $ & $ 361_{-51}^{+56}$ &  $0.177_{-0.022}^{+0.026}$ & $0.374_{-0.065}^{+0.058}$ & $0.216_{-0.055}^{+0.054}$   &  $5.9_{-0.7}^{+0.9}$ &  $9.2_{-1.6}^{+1.4}$ & $5.3_{-1.4}^{+1.3}$  \\
PJ338+29    &$6.6668\pm0.0004$ & $<0.057       $ & $ 399_{-64}^{+76}$ &  $<0.111$		   & $0.221_{-0.054}^{+0.050}$ & $<0.135$		     &  $<4.2$  	    &  $6.1_{-1.5}^{+1.4}$ & $<3.6$		  \\
PJ359--06   &$6.1726\pm0.0004$ & $<0.050       $ & $ 358_{-50}^{+62}$ &  $<0.132$		   & $0.235_{-0.068}^{+0.060}$ & $<0.153$		     &  $<4.5$  	    &  $5.8_{-1.7}^{+1.5}$ & $<3.9$		  \\
\hline			        		        										  								      
Stack       &  &  $0.153\pm0.011$ & $ 323_{-24}^{+24}$ &  $0.172_{-0.014}^{+0.015}$ & $0.292_{-0.024}^{+0.027}$ & $0.105_{-0.028}^{+0.021}$   &  $5.6_{-0.4}^{+0.5}$ &  $7.0_{-0.6}^{+0.6}$ & $2.5_{-0.7}^{+0.5}$  \\
\hline
\end{tabular}
\end{center}
\end{table*}

\subsection{Observations and data reduction}\label{sec_noema}

The dataset consists of the collection of a main survey and a number of smaller programs conducted at NOEMA and previously at PdBI, for a total observing time of $\sim70$\,h on source (8 antennas equivalent). The half power beam width of NOEMA's 15\,m antennas is $\approx 48.6''$ at 100\,GHz. Observations were mostly carried out under average or poor weather conditions with precipitable water vapour columns typically $\gsim$5\,mm and system temperatures of 80--200\,K with the array in compact (C or D) configuration. We processed the data using \textsf{clic} from the \textsf{GILDAS}\footnote{https://www.iram.fr/IRAMFR/GILDAS/} suite. Table~\ref{tab_obs} provides a summary of our observations.

J0338+0021 was observed on August 1, and September 28, 2008 (program ID: S052), using the narrow-band 2\,mm receiver, and again in July 2013 (program ID: X04D) using WideX. The sources 3C454.3 and MWC349 were observed for bandpass and flux calibration. We observed J0406+066 as amplitude and phase calibrator. The final cube includes 7200 visibilities, corresponding to 9.0\,h (5-antenna equivalent).

J2054--0005 was observed in October--November 2013 (program ID: X04D). We used the WideX correlator and tuned the receiver to secure the \Ci{} and CO(7--6) lines in a single setup. The sources 3C454.3 and MWC349 were observed for bandpass and flux calibration. For the amplitude and phase calibration, we observed 2059+034. The final cube includes 16881 visibilities, corresponding to 21.10\,h (5-antenna equivalent).

PJ036+03 was observed in multiple short tracks in June--July 2015 (program ID: S15DA) and June--July 2017 (program ID: S17CD), in compact array configuration (5D / 6D in 2015; 7D / 8D in 2017) using the WideX correlator. The receiver was tuned to encompass the two CO(6--5) and CO(7--6) lines in two independent frequency settings. The calibration sources include the blazar 3C454.3 for bandpass calibration, the source MWC349 for absolute flux calibration, and the quasar 0215+015 for amplitude and phase calibration. The final cubes consist of a total of 22440 visibilities (13.33\,h on source, assuming 7 antennas) on the CO(7--6) line, and 22317 visibilities (13.28\,h on source, assuming 7 antennas) on CO(6--5).

The quasar PJ338+29 was observed in July 2018 (program ID: S18DM). The expanded bandwidth offered by the PolyFix correlator enables to encompass the CO(6--5) and CO(7--6) lines in our targets using a single frequency setting. The blazar 3C454.3 served as bandpass calibrator, MWC349 as absolute flux calibrator, and the quasars 2234+282 and J2217+243 as amplitude and phase calibrators. The final cubes comprises a total of 8670 visibilities (3.87\,h on source, 8-antenna equivalent).

We observed the quasar J2219+0102 as part of two programs, one covering the CO(5-4) and CO(6--5) lines (August--September 2018, program ID: S18DM), and one covering the CO(6--5), CO(7--6) and \Ci{} lines (June 2019, program ID: S19DM). The bright sources 3C454.3 and MWC349 acted as bandpass and flux calibrators, while the quasars 2216-038 and 2223-052 were observed as amplitude and phase calibrators. We reduced the two settings independently, and then merged the visibilities in the overlapping frequency range. The 2018 and 2019 observations comprise 13380 and 8640 visibilities, respectively, corresponding to 5.97 and 3.86\,h on source (8-antenna equivalent).

We observed the remaining targets between June 2019 and January 2020 (program ID: S19DM). The observations of PJ159-02 relied on 0851+202, LKHA101, and J1028-0236 as bandpass, flux, and phase/amplitude calibrators respectively. The final cube consists of 7170 visibilities (2.0\,h on source, 10-antenna equivalent). Observations of J1319+0950 were performed in excellent weather conditions (precipitable water vapour $<2$\,mm, system temperature $T_{\rm sys}$=55-70\,K) and used on the calibrators 3C273 (bandpass), MWC349 (flux), and 1307+121 (phase, amplitude, pointing, focus). The final cubes consists of 8310 visibilities (2.88\,h on source, 9-antenna equivalent). 
Observations of the quasar J1048-0109 were performed using 3C84, 3C273, and LKHA101 as bandpass and flux calibrators, and 1055+018 as phase and amplitude calibrator. The final cube consists of 10170 visibilities (3.53\,h on source, 9-antenna equivalent). We calibrated the observations of the quasar J1104+2134 using MWC349, 0851+202, 3C273, 0923+392 for flux and bandpass calibration; and 1040+244 for phase and amplitude. The final cube consists of 12660 visibilities (4.40\,h on source, 9-antenna equivalent). Finally, the observations of PJ359-06 capitalized on 3C454.3, MWC349, and 0003-066 for bandpass, flux, and amplitude/phase calibration. The final cube comprises 3.46\,h on source (9-antenna equivalent).

\begin{figure}	  
\begin{center}
\includegraphics[width=0.49\textwidth]{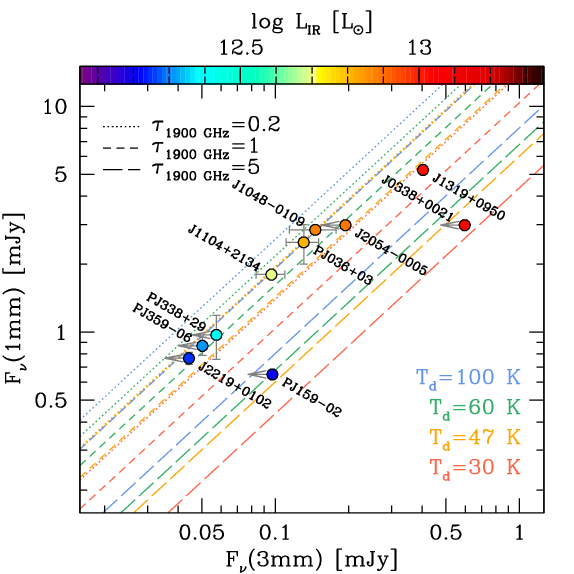}\\
\end{center}
\caption{Measured continuum flux densities at 3\,mm (at $\nu\approx\nu_{\rm CO(7-6)}$) and 1\,mm (at $\nu\approx \nu_{\rm [CII]}$). The expected flux density ratios for modified black body templates with different temperatures $T_{\rm d}$ and opacity at the \Cii{} frequency, corrected for the CMB for a source at redshift $z=6$, are shown for comparison. Our measurements, color-coded by the IR luminosity of the quasars, are generally consistent with $T_{\rm d}$=47\,K and $\tau_{\rm 1900\,GHz}$=0.2, or $T_{\rm d}$=60\,K and $\tau_{\rm 1900\,GHz}$=1. The former template is adopted in our analysis when computing IR luminosities and dust masses.}\label{fig_1mm_3mm}
\end{figure}


\subsection{Imaging}\label{sec_image}

We imaged the cubes using the \textsf{GILDAS} suite \textsf{mapping}. Natural weighting was adopted in all the cases. The beam FWHM ranges between $3''$ and $7''$ (see Table~\ref{tab_obs}). We resampled the spectral axis in 50\,\kms{} wide channels (computed at the CO(7--6) frequency). In creating the continuum images, we first masked out the channels where strong emission lines are found, based on visual inspection of the spectra. When no obvious lines were detected, we masked out a range of $\sim 600$\,\kms{} around the expected frequency of the lines. We created continuum images from the lower and upper side bands independently using the unmasked channels, and use them to subtract from the cubes using the task \textsf{uv\_subtract}. Finally, we created continuum--subtracted line maps using the channels within the line full width at zero intensity. In the absence of an obvious line detection, we used a 6 channels (300\,\kms{}) window around the expected frequency of the transition, based on the \Cii{} redshift. For all the maps, we defined cleaning boxes around the quasar, and cleaned down to 1.5-$\sigma$ per channel using Hogbom cleaning (via the task \textsf{clean}). Finally, we extracted spectra in single-pixel extraction at the formal optical/near-infrared coordinates of the targets (i.e., we did not optimize the signal extraction based on the observed maps). All astrometric uncertainties are negligible with respect to the beam size of the presented observations. Figs.~\ref{fig_spc1}--\ref{fig_spc3} show the extracted spectra and the maps of all the sources in our sample.

\section{Results}\label{sec_results}

In the following, we present our results on the spectral analysis of the new CO(7--6) + \Ci{} and CO(6--5) observations (sec.~\ref{sec_fits}), we infer molecular gas mass estimates (sec.~\ref{sec_masses}), and constrain the properties of the ISM based on the available lines (sec.~\ref{sec_ism}).

\subsection{Spectral analysis}\label{sec_fits}

We fit the 3\,mm spectra from our sample and the 1\,mm photometric measurements from the literature (see Table~\ref{tab_sample}) with a Gaussian curve for each line: CO(6--5), CO(7--6), and \Ci{}; and a dust continuum described as:
\begin{equation}\label{eq_cont}
S_{\nu, \rm obs}=\frac{\Omega}{(1+z)^3}\,[B_\nu(T_{\rm d})-B_\nu(T_{\rm CMB})]\,[1-\exp(-\tau_\nu)]
\end{equation}
where $\Omega$ is the source angular size in steradiants; $z$ is the source redshift; $B_\nu(T)$ is the black body emissivity spectrum; $T_{\rm d}$ is the dust temperature; $T_{\rm CMB}$ is the CMB temperature at the source redshift, $T_{\rm CMB}=2.725\,(1+z)$\,K. Finally, $\tau_\nu$ is the optical depth:
\begin{equation}\label{eq_tau}
\tau_\nu = \kappa_{\rm d}(\nu)\,\Sigma_{\rm d} \approx \kappa_{\rm d}(\nu)\,\frac{M_{\rm d}}{\Omega\,D_{\rm A}^2} = \kappa_{\rm d}(\nu)\,\frac{N_{\rm H2}\,m_{\rm H2}}{\delta_{\rm g/d}}
\end{equation}
where $\kappa_{\rm d}(\nu)$ is the frequency--dependent dust absorption coefficient; $\Sigma_{\rm d}$ is the dust surface density; $M_{\rm d}$ is the total dust mass within the angular size defined by $\Omega$; $D_{\rm A}$ is the angular diameter distance, $D_{\rm A}=D_{\rm L} / (1+z)^2$; $N_{\rm H2}$ is the column density of molecular gas; $m_{\rm H2}$ is the mass of the H$_2$ molecule; and $\delta_{\rm g/d}$ is the gas--to--dust mass ratio. We interpolate the $\kappa_{\rm d}(\nu)$ values reported in Table 5 of \citet{draine03} throughout the IR regime, and extrapolate assuming a power-law with slope $\beta$=1.9 at $\nu<350$\,GHz \citep[see also][]{dacunha21}. At the CO(7--6) frequency, eq.~\ref{eq_tau} implies that $\tau_\nu\sim 1$ for a molecular gas column density of $\sim 1.8\times 10^{25}$\,cm$^{-2}$ (assuming $\delta_{\rm g/d}$=100, see, e.g., \citealt{berta16}). Such high column densities have been observed in the nuclei of some ULIRGs \citep[e.g., Arp\,220; see][]{scoville17b}, but are not expected to apply to the galaxy--averaged regimes sampled in our spatially--unresolved observations (see also sec.~\ref{sec_ism}). At \Cii{} frequencies, the same optical depth is reached at $\approx 6.3\times$ lower column densities. 

We perform the spectral fits using our custom Monte Carlo Markov Chain algorithm \textsf{smc}. We fit the redshift of the quasars, informed by the \Cii{} redshift; the width of the emission lines; the integrated fluxes of the lines; the dust continuum temperature, $T_{\rm d}$; and the dust mass within the beam, $M_{\rm d}$. Due to the modest signal-to-noise ratio of some of the lines in our survey, we impose that all of the fitted CO and \Ci{} lines of a given source have the same velocity width. As priors, we adopt Maxwellian distributions for the line width and the dust temperature (using 300\,\kms{} and 50\,K as scale lengths of the distributions); and loose Gaussian priors for the line fluxes and the logarithm of the dust mass. In equations~\ref{eq_cont}--\ref{eq_tau}, we assume $\Omega$=$\Omega_{\rm b} f$, where $\Omega_{\rm b}$ is the resolution element in steradiants, and $f$ is the (unconstrained) filling factor. Operatively, we assume $f$=1 in the fits, and stress that, in the optically--thin scenario, $\Omega$ has no impact on $S_\nu$ and $M_{\rm d}$ as it disappears from eq.~\ref{eq_cont}. The best fit and 1-$\sigma$ errors on the fitted parameters are derived from the 50\%, 14\%, and 86\% quantiles of the marginalized posterior distributions. The results of the fits are listed in Table~\ref{tab_fits}. In the following, we treat as upper limits measurements of line fluxes and continuum flux densities for which the posterior flux estimates are below their upper 3-$\sigma$ confidence level. All of the quasars are detected in the CO(7--6) emission line. Conversely, only 6/10 and 6/8 of the targeted quasars are detected in their \Ci{} and CO(6--5) emission, respectively.

In Fig.~\ref{fig_1mm_3mm}, we plot the continuum flux density at 1\,mm (at the rest frequency of \Cii{}, as reported in the literature; see Table~\ref{tab_sample} for references) as a function of the observed flux densities at 3\,mm [at the reference frequency of CO(7--6), from this work]. We find a typical 1\,mm/3\,mm flux density ratio of $\sim 19$. Fig.~\ref{fig_1mm_3mm} also shows the flux densities expected under different assumptions of $T_{\rm d}$ and $\tau_{\rm 1900\,GHz}$, the optical depth $\tau_{\nu}$ computed at the \Cii{} frequency, for a quasar at $z$=6, based on Eq.~\ref{eq_tau}. The observed ratios favor relatively low optical depth values, $\tau_{\rm 1900\,GHz}\lsim 1$ against the optically-thick scenario ($\tau_{\rm 1900\,GHz}\gsim 5$) for the quasars in our sample, unless extremely high temperatures ($T_{\rm d}\gg 100$\,K) are invoked. This suggests that the bulk of the dust is optically thin at 3\,mm in these galaxy-wide observations, in contrast with what has been observed at the very center of some quasar host galaxies \citep[e.g.,][]{walter22}.

For most of the quasars in our sample, only the two photometric points shown in Fig.~\ref{fig_1mm_3mm} are available in the rest-frame far-infrared, thus making our estimates of the IR luminosities quite uncertain. In particular, the lack of constraints on the dust spectral energy distribution at wavelengths shorter than the expected peak hinders any informative constraints on the dust temperature, $T_{\rm d}$. E.g., the observed $F_{\nu}$(1mm)/$F_{\nu}$(3mm) ratios in our sample are equally fit with ($T_{\rm d}$,$\tau_{\rm 1900 GHz}$)=(60\,K, 1) or (47\,K, 0.2). With this caveat in mind, in the remainder of the analysis we infer IR luminosities following Eq.~\ref{eq_tau} under the assumption of $\tau_{\rm 1900\,GHz}=0.2$ and $T_{\rm d}$=47\,K, scaled to match the observed 1\,mm flux density. These values are in line with the dust temperature reported for other high--$z$ quasars that have been extensively studied in their dust spectral energy distribution \citep[e.g.,][]{beelen06,drouart14,leipski14,meyer22}. For the quasars J0338+0021, J2054--0005, and J1319+0950, we instead adopt the values published in \citet{wang13} and \citet{leipski14}, based on the fits of their Spectral Energy Distributions. In all cases, the IR luminosities are computed by integrating the dust continuum between 8--1000 $\mu$m.

We convert line fluxes into luminosities, following, e.g., \citet{carilli13}:
\begin{equation}\label{eq_lum1}
\frac{L'}{\rm K\,km\,s^{-1}\,pc^2} = \frac{3.25\times 10^7}{1+z} \frac{F_{\rm line}}{\rm Jy\,km\,s^{-1}}\,\left(\frac{\nu}{\rm GHz}\right)^{-2}\,\left(\frac{D_{\rm L}}{\rm Mpc}\right)^2
\end{equation}
\begin{equation}\label{eq_lum2}
\frac{L}{\rm L_\odot} = \frac{1.04\times 10^{-3}}{1+z} \frac{F_{\rm line}}{\rm Jy\,km\,s^{-1}}\,\frac{\nu}{\rm GHz}\,\left(\frac{D_{\rm L}}{\rm Mpc}\right)^2
\end{equation}
where $F_{\rm line}$ is the line integrated flux. 

\begin{figure}
\begin{center}
\includegraphics[width=0.49\textwidth]{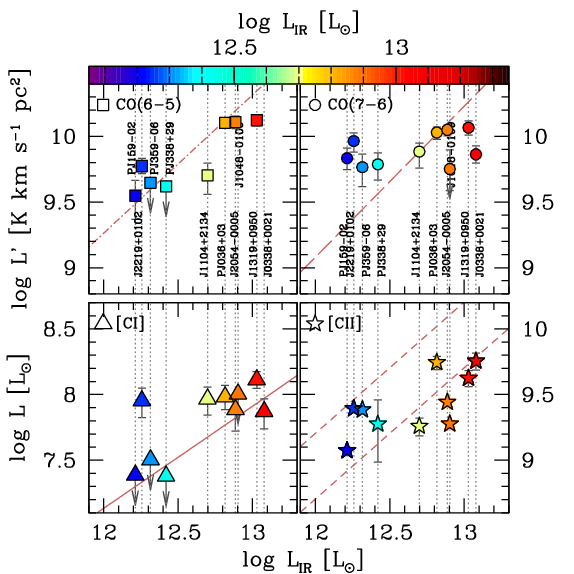}\\
\end{center}
\caption{The luminosity of CO(6--5), CO(7--6), \Ci{}, and \Cii{}, as a function of IR luminosity, for the quasars in our sample. For reference, we also show the CO line--IR luminosity relations from \citet{greve14} (dashed--dotted and long--dashed lines); the \Ci{}--$L_{\rm IR}$ relation from \citet{valentino20} (solid line); and the two average values of the \Cii{}/IR luminosity ratio for low-- and high--$\Sigma_{\rm IR}$ galaxies from \citet{herreracamus18} (short dashed lines). The targeted quasars span a factor $\sim 10$ in IR luminosity but only a factor $\gsim 3$ in each line luminosity.}\label{fig_co_fir}
\end{figure}

In Fig.~\ref{fig_co_fir}, we compare the luminosity of the lines under examination with the infrared luminosity. Our sample spans a factor $\sim10$ in IR luminosity, but only a factor of $\lsim 3$ in the luminosity of any of the detected lines. Both the measured CO(6--5) and CO(7--6) luminosities are in line with empirical scaling relations based on local ULIRGs \citep[see, e.g.,][]{greve14}. This, together with the widespread detections of CO(7--6), and the similar luminosities found for CO(6--5) and CO(7--6), suggests that the CO excitation is relatively high up to $J_{\rm up}\approx 7$. We measure \Ci{} luminosities of $(3-10)\times10^7$\,\Lsun{}, as expected based on the empirical \Ci{}--IR relation from \citet{valentino20}. Finally, the \Cii{}/IR luminosity ratios in our sample span a factor $\sim6$, from $1.4\times10^{-3}$ (J2219+0102) to $2.4\times10^{-4}$ (J2054--0005), with lower \Cii{}/IR ratios typically observed for quasars with $L_{\rm IR}>3\times10^{12}$\,\Lsun{} \citep[see, e.g.,][]{stacey10,gullberg15}. For comparison, the observed ratios in low/high surface density of star formation rate regimes as computed in \citet{herreracamus18} are \Cii{}/IR$\approx$1.6$\times10^{-3}$ and \Cii{}/IR$\approx$5$\times10^{-4}$, respectively (assuming the SFR--IR conversion from \citealt{kennicutt12}).

\begin{figure}
\begin{center}
\includegraphics[width=0.49\textwidth]{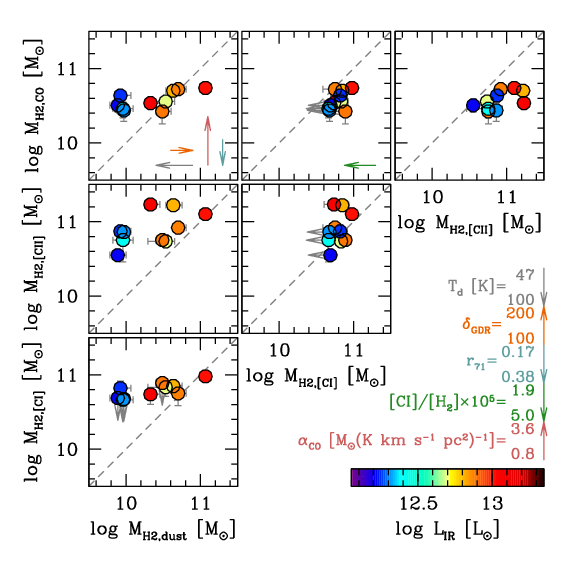}\\
\end{center}
\caption{Comparison between the mass estimates inferred from CO ($M_{\rm H2,CO}$), \Ci{} ($M_{\rm H2,[CI]}$), \Cii{} ($M_{\rm H2,[CII]}$) and dust continuum ($M_{\rm H2,dust}$), as described in the text. The one--to--one relation is shown as a dashed line. The different estimators lead to generally consistent results overall, with $M_{\rm H2}=10^{10-11}$\,\Msun{}, although different estimators lead to a different spread of values within our sample. The \Cii{}--based estimates are systematically higher than the ones based on other tracers. The arrows highlight how the data points would move if some parameters would change from their fiducial values to others commonly used in the literature.}\label{fig_masses}
\end{figure}

Capitalizing on the self-similarity of the line luminosities within our sample, we compute the average spectrum by shifting all of the 3\,mm spectra to $z=6$, and averaging the observed flux densities with weights set by the channel variance. The noise spectrum is derived as the sum of the inverse variance of the spectra of each source. We do not rescale the spectra based on the source luminosity. The stacked spectrum is then fitted in the similar fashion as individual sources. The stacked spectrum and the fitted values are reported in Fig.~\ref{fig_spc3} and Table~\ref{tab_fits}. In the following, when comparing the luminosity of the stacked lines with the one of \Cii{}, for the stack we adopt the median luminosity of \Cii{} in our sample, $L_{\rm [CII]}=2.5\times 10^9$\,\Lsun{}.

\subsection{Molecular gas masses}\label{sec_masses}

Here we infer molecular mass estimates via four recipes, based on the dust continuum emission and different molecular gas tracers. This allows us to test the consistency of the various mass estimators and to study the impact of working assumptions. 

First, we derive the dust mass from the dust continuum emission. As discussed in the previous section, $\tau_\nu\ll 1$ at 3\,mm in our galaxy--scale observations. In this regime, the dust is optically--thin and eq.~\ref{eq_cont} yields a nearly linear dependence of the observed flux density on $\tau_\nu$, and hence, on the dust mass. In the following, we adopt the template used for the estimate of the IR luminosiy ($T_{\rm d}$=47\,K, $\tau_{\rm 1900\,GHz}$=0.2). Because the optical depth is fixed, eq.~\ref{eq_cont} implies a constraint on the size of the emitting region, $\Omega$, which thus sets the dust mass via eq.~\ref{eq_tau}. In this framework, an observed 3\,mm flux density of 0.1\,mJy for a source at $z$=6 corresponds to a size of $0.097$\,arcsec$^2$ ($\sim 3.2$\,kpc$^2$), and a dust mass of $M_{\rm d}=2.8\times10^8$\,\Msun{}. Our CO(7--6) observations have a median beam area of 16.8 arcsec$^2$, yielding a typical filling factor $f$=$5.8\times10^{-3}$. If instead we leave $\tau_{\nu}$ free, the dependence on the angular size is lost in the optically--thin regime, and the flux density is simply a function of temperature and dust mass. As discussed in sec.~\ref{sec_fits}, however, we cannot constrain the dust temperature in the majority of our sample. If we refer to the best--fit values from our continuum fits, $T_{\rm d}$=30--110\,K, the inferred dust masses differ from the fiducial ones by an average factor $\sim 1.6$. 
%
%
%
%
We convert the dust mass into the associated molecular gas mass via a gas--to--dust ratio, $\delta_{\rm g/d}$=100 \citep[see, e.g.,][]{bolatto13,sandstrom13,genzel15,berta16}. For comparison, the analysis by \citet{dunne21} led to a mean value of $\delta_{\rm g/d}=128\pm16$, in broad agreement with the round value assumed here. The inferred molecular gas masses are listed in Table~\ref{tab_masses}.

Molecular gas mass estimates in high--$z$ galaxies typically rely on CO observations \citep[see, e.g., reviews in][]{carilli13,combes18,tacconi20,hodge20}. While low--J transitions should be preferred, as they are less sensitive to uncertainties on the CO excitation, the molecular mass estimates in $z>6$ quasars tend to rely on intermediate (J$_{\rm up}$=5--7) transitions \citep[e.g.,][]{wang10,venemans17a,yang19} which appear to be at the peak of the CO Spectral Line Energy Distribution for these quasars \citep[e.g.,][]{li20}. In this work, we pin our estimates to the CO(7--6) emission\footnote{CO(7--6) is preferred to CO(6--5) as it is available for all of the sources, and in our dataset it usually displays higher signal-to-noise.}. Molecular gas mass is derived from CO as:
\begin{equation}\label{eq_mco}
M_{\rm H2}=\alpha_{\rm CO}\,r_{71}^{-1}\,L'_{\rm CO(7-6)}
\end{equation}
where we adopt a CO--to--H$_2$ conversion factor $\alpha_{\rm CO}$=0.8\,\Msun{}\,(\Kkmspc)$^{-1}$ typical for ULIRG and quasars \citep[see discussions in][]{bolatto13,carilli13}, and $r_{71}$ is the CO(7--6)/CO(1--0) luminosity ratio, that we assume to be $r_{71}$=$0.17$ based on the high--$z$ CO excitation template in \citet{boogaard20}. For comparison, the well-studied quasar J1148+5251 shows $r_{71}>0.12$, and $r_{73}=0.58$, suggesting very high CO excitation \citep{bertoldi03,riechers09}. Table~\ref{tab_masses} lists the inferred values. 

A third path to molecular gas estimates capitalizes on the neutral carbon emission, which arises from the outer layers of molecular clouds as well as the diffuse molecular gas in the interstellar medium (see theoretical predictions in, e.g., \citealt{glover16,clark19,bisbas19}; Galactic observations in, e.g., \citealt{frerking89,kramer04,cubick08,beuther14}; and searches at high redshift by, e.g., \citealt{walter11,alaghband13,bothwell17,popping17,papadopoulos18,valentino18}). In the optically-thin assumption, following a three--level model of the carbon emission, the mass of neutral carbon can be inferred from the observed line luminosity using:
\begin{equation}\label{eq_mci}
\frac{M_{\rm CI}}{\rm M_\odot} = \frac{4.556}{10^{4}}\,\frac{Q_{\rm ex}}{5}\,\exp(62.5/T_{\rm ex})\,\frac{L'_{\rm [CI]2-1}}{\rm K\,km\,s^{-1}\,pc^2}
\end{equation}
where $Q_{\rm ex}=1+3\,\exp(-23.6/T_{\rm ex})+5\,\exp(-62.5/T_{\rm ex})$ is the partition function and $T_{\rm ex}$ is the excitation temperature in Kelvin \citep[see][]{weiss03,weiss05}. We assume that the carbon emission is in thermal equilibrium with the dust, and thus adopt $T_{\rm ex}=47$\,K as measured on average in high--$z$ quasars \citep{beelen06,leipski14}. The mass estimate in eq.~\ref{eq_mci} is not very sensitive to $T_{\rm ex}$: any temperature in the range $T_{\rm ex}$=30--100\,K yields a mass estimate within $\sim 0.15$ dex from our fiducial value. The resulting masses can be converted into H$_2$ masses through the neutral carbon to H$_2$ abundance, [C]/[H$_2$], and accounting for the mass ratio of carbon atoms and H$_2$ molecules. Following \citet{boogaard20}, here we adopt a carbon abundance value of $(1.9\pm0.4)\times 10^{-5}$ \citep[under the assumption that CO is in thermal equilibrium; see][]{bothwell17,boogaard20}. This value is consistent with the [C]/[H$_2$]=$(1.6\pm0.1)\times10^{-5}$ estimate by \citet{dunne21} for a sample of $z=0.35$ galaxies.

\begin{table*}
\caption{Inferred quantities from the observed line and continuum luminosities. (1) Target name. (2--5) Molecular gas mass estimates from dust continuum ($M_{\rm H2,dust}$), CO ($M_{\rm H2,CO}$), \Ci{} ($M_{\rm H2,[CI]}$), and \Cii{} ($M_{\rm H2,[CII]}$), derived as described in the text. (6--8) Observed line luminosity ratios. }\label{tab_masses}
\begin{center}
\begin{tabular}{c|cccc|ccc}
\hline
Target      & $M_{\rm H2,dust}$       & $M_{\rm H2,CO}$       & $M_{\rm H2,[CII]}$    & $M_{\rm H2,[CI]}$    & log \Cii{}/\Ci{}           & log \Cii{}/CO(7--6)        & log \Ci{}/CO(7--6)          \\
            & [$10^{10}$\,M$_\odot$]  & [$10^{10}$\,M$_\odot$]& [$10^{10}$\,M$_\odot$]&[$10^{10}$\,M$_\odot$]&                            &                           &                            \\
 (1)        & (2)                     & (3)                   & (4)                   & (5)                  & (6)                        & (7)                       & (8)                        \\
\hline
PJ036+03    & $ 4.33_{-0.35}^{+0.53}$ & $ 5.0_{-0.6}^{+0.5}$  & $16.66_{-1.9}^{+1.9}$ & $ 7.1_{-1.4}^{+1.6}$ & $1.762_{-0.106}^{+0.105}$  & $1.490_{-0.069}^{+0.069}$ & $-0.273_{-0.107}^{+0.100}$ \\
J0338+0021  & $ 2.14_{-0.08}^{+0.21}$ & $ 3.4_{-0.5}^{+0.6}$  & $17.07_{-2.5}^{+2.5}$ & $ 5.5_{-1.4}^{+1.4}$ & $1.884_{-0.123}^{+0.142}$  & $1.667_{-0.098}^{+0.087}$ & $-0.217_{-0.153}^{+0.114}$ \\
PJ159--02   & $ 0.78_{-0.07}^{+0.09}$ & $ 3.2_{-0.6}^{+0.6}$  & $ 3.55_{-0.2}^{+0.2}$ &      $<4.9$          & $>1.252$                   & $1.015_{-0.083}^{+0.090}$ & $<-0.237$                  \\
J1048--0109 & $ 5.03_{-0.95}^{+1.92}$ & $ 5.3_{-0.5}^{+0.5}$  & $ 8.32_{-0.2}^{+0.2}$ & $ 5.7_{-1.8}^{+1.6}$ & $1.558_{-0.108}^{+0.164}$  & $1.168_{-0.039}^{+0.046}$ & $-0.390_{-0.170}^{+0.114}$ \\
J1104+2134  & $ 3.42_{-1.99}^{+1.50}$ & $ 3.6_{-0.9}^{+0.6}$  & $ 5.45_{-0.8}^{+0.8}$ & $ 6.8_{-1.7}^{+1.6}$ & $1.293_{-0.121}^{+0.134}$  & $1.150_{-0.101}^{+0.137}$ & $-0.143_{-0.144}^{+0.149}$ \\
J1319+0950  &  $11.7_{-8.6}^{+10.8}$  & $ 5.5_{-0.7}^{+0.7}$  & $12.66_{-1.8}^{+1.8}$ & $ 9.6_{-1.3}^{+1.5}$ & $1.513_{-0.092}^{+0.084}$  & $1.334_{-0.085}^{+0.078}$ & $-0.179_{-0.086}^{+0.082}$ \\
J2054--0005 & $ 3.09_{-0.19}^{+0.32}$ & $ 2.7_{-0.9}^{+1.1}$  & $ 5.67_{-0.3}^{+0.3}$ &      $<7.8$          & $>1.253$                   & $1.300_{-0.149}^{+0.167}$ & $<0.046$                   \\
J2219+0102  & $ 0.84_{-0.05}^{+0.13}$ & $ 4.3_{-0.8}^{+0.7}$  & $ 7.44_{-0.4}^{+0.4}$ & $ 6.6_{-1.7}^{+1.6}$ & $1.441_{-0.100}^{+0.129}$  & $1.204_{-0.068}^{+0.086}$ & $-0.236_{-0.148}^{+0.123}$ \\
PJ338+29    & $ 0.91_{-0.09}^{+0.13}$ & $ 2.9_{-0.7}^{+0.7}$  & $ 5.66_{-2.8}^{+3.0}$ &      $<4.6$          & $>1.479$                   & $1.263_{-0.320}^{+0.209}$ & $<-0.216$                  \\
PJ359--06   & $ 0.94_{-0.08}^{+0.08}$ & $ 2.7_{-0.8}^{+0.7}$  & $ 7.27_{-0.5}^{+0.4}$ &      $<4.7$          & $>1.575$                   & $1.394_{-0.105}^{+0.149}$ & $<-0.181$                  \\
\hline
Stack\tablefootmark{a} &              &                       &                       &                      & $1.772_{-0.138}^{+0.155}$  & $1.327_{-0.114}^{+0.091}$ & $-0.445_{-0.145}^{+0.086}$ \\
\hline
\end{tabular}
\end{center}
\tablefoot{
\tablefoottext{a}{For the line ratios of the stacked spectrum, we adopt the average value $L_{\rm [CII]}=2.5\times10^9$\,\Lsun{} based on the median value of the literature measurements in} Table~\ref{tab_sample}.
}
\end{table*}

Finally, recent work on intermediate redshift main sequence galaxies led to a definition of a molecular mass estimator based on \Cii{} \citep{zanella18}:
\begin{equation}\label{eq_mcii}
M_{\rm H2,[CII]} = \alpha_{\rm [CII]}\,L_{\rm [CII]},
\end{equation}
where $L_{\rm [CII]}$ is the \Cii{} luminosity in solar units, and $\alpha_{\rm [CII]}\sim 30$\,M$_\odot$\,L$_\odot^{-1}$ is a scaling factor calibrated on a collection of main sequence and starburst galaxies. \citet{venemans17b} provide a first principle estimate of the ionized carbon mass in analogy to eq.~\ref{eq_mci}. Assuming $T_{\rm ex}=47$\,K, this formula is consistent with Eq.~\ref{eq_mcii} if one assumes a [C$^+$]/[H$_2$] abundance of $3.4\times10^{-5}$, i.e., about 1.8 times higher than the [C]/[H$_2$] abundance from \citet{boogaard20} but still in line with other carbon abundance values reported in the literature \citep[e.g.,][]{weiss05}. On the other hand, \citet{sommovigo21} find a wide range of $\alpha_{\rm [CII]}$=10--1000 for a sample of local galaxies, depending on the surface density of SFR and on the depletion time.

Figure~\ref{fig_masses} and Table~\ref{tab_masses} compare the molecular gas masses inferred with these four methods. All the $M_{\rm H2}$ estimates are in broad agreement, with typical values of $10^{10-11}$\,\Msun{}, although the \Cii{}--based values show a systematic excess by $\sim 0.5$\,dex. Noticeably, \citet{neeleman21} find a similar excess for \Cii{}--based molecular gas mass measurements, when comparing to the dynamical masses obtained from high--resolution \Cii{}  maps of quasars at similar redshifts. The dust--based estimates show the largest scatter in our work, as they spread over $\sim 1$ dex; on the contrary, the CO--based estimates display a much smaller range ($<0.5$\,dex). The plot also highlights the impact of our operative assumptions: Our $M_{\rm H2}$ estimates depend linearly on $\alpha_{\rm CO}$,  $\delta_{\rm g/d}$, and $\alpha_{\rm [CII]}$, and anti-linearly on $r_{71}$, [C]/[H$_2$], and (to first order) $T_{\rm d}$. Different assumptions for these parameters could mitigate or increase discrepancies among these mass estimates. E.g., by adopting a higher CO--to--H$_2$ value, $\alpha_{\rm CO}=3.6$\,M$_\odot$\,(\Kkmspc)$^{-1}$ \citep[see, e.g.,][]{daddi10}, the CO--based estimates of $M_{\rm H2}$ would increase by a factor $4.5$, thus narrowing the gap with respect to the \Cii{}--based ones, but at the same time separating them from the \Ci{}-- and dust--based measurements. On the other hand, if we adopt a higher CO--excitation, e.g., $r_{71}=0.38$ from the quasar template in \citet{carilli13} instead of the fiducial value $r_{71}=0.17$ from \citet{boogaard20}, we would reduce the CO--based dust mass estimates, thus widening the gap with respect to the \Cii{}--based estimates.  Similarly, assuming a higher carbon abundance, [C]/[H$_2$]=$5\times10^{-5}$ as estimated by \citet{weiss05} in a sample of $z\sim2.5$ sub-mm galaxies, would yield a decrease in the \Ci{}--based mass of molecular gas. Along the same lines, one could adopt the method proposed by \citet{sommovigo21} to infer $\alpha_{\rm [CII]}$ and $T_{\rm d}$ based on the surface brightness of \Cii{} for those quasars in our sample that have already been observed at high angular resolution in \Cii{} \citep[PJ036+03, J1048--0109, J1319+0950, J2054--0005, and PJ359--06; see][]{venemans20}. The extremely high SFR densities reached in these sources lead to very low $\alpha_{\rm [CII]}\sim 1$ values and correspondingly to slightly higher dust temperatures, $T_{\rm d}$=40--100\,K. The higher $T_{\rm d}$ would yield slightly lower ($<0.2$\,dex on average) $M_{\rm H2,dust}$ estimates. However, the $M_{\rm H2,[CII]}$ estimates would drop by a factor $\sim30$ based on this method, making them inconsistent with all other estimates. This discrepancy is likely due to the implicit wild extrapolation of local scaling relations (such as the Kennicutt--Schmidt star-formation law and the empirical SFR--$L_{\rm [CII]}$ relation) towards regimes where other astrophysical mechanisms may be at play (for instance, \Cii{} might be suppressed by thermalization, changes in the photoelectric efficiency of dust grains, optical depth, and other processes; see e.g. \citealt{sutter21}).

While a deeper understanding of the physical properties of the line and dust emitting ISM is required in order to directly constrain these unknown quantities, our analysis demonstrates that the fiducial values lead to consistent molecular gas mass estimates (within a factor $\lsim 3$).

\begin{figure*}
\begin{center}
\includegraphics[width=0.49\textwidth]{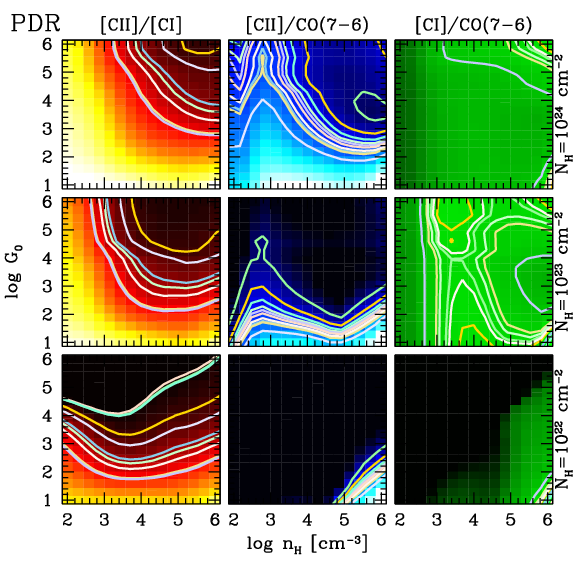}
\includegraphics[width=0.49\textwidth]{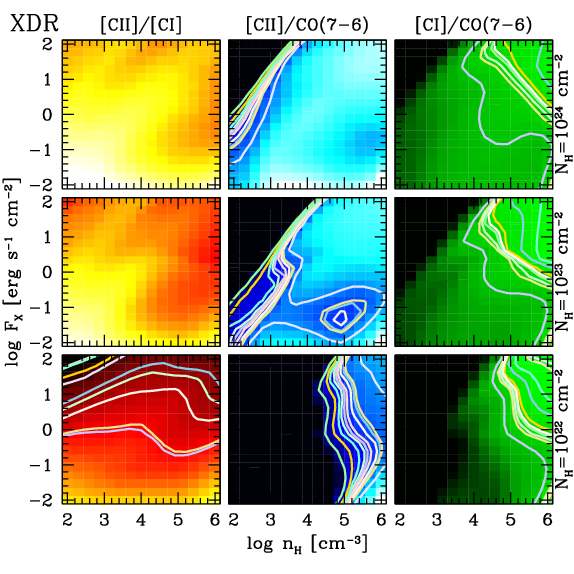}\\
\includegraphics[width=0.49\textwidth]{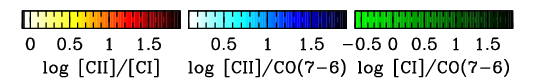}
\includegraphics[width=0.49\textwidth]{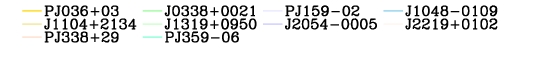}\\
\end{center}	 
\caption{Luminosity line ratios predicted in the context of PDR / XDR models, as a function of column density, gas density, and strength of the incident radiation field (see text for details). The contours show the ratios measured in each source in our sample, color coded as shown in the legend. XDR models can reproduce the observed high \Cii{}/\Ci{} ratios only at low column densities, $N_{\rm H}\lsim 10^{22}$\,cm$^{-2}$, whereas PDR models are compatible with the observed constraints over a wider range of parameters. The relatively low \Cii{}/CO(7--6) ratio observed in the targeted quasars is suggestive of typically large column densities, $N_{\rm H}\gsim 10^{23}$\,cm$^{-2}$ for both PDR and XDR environments. Finally, the low observed \Ci{}/CO(7--6) ratios point to a high density $n_{\rm H}\gsim 10^3$ ($10^5$)\,cm$^{-3}$ in PDRs (XDRs). Overall, a dense ($n_{\rm H}>10^3$\,cm$^{-3}$) gas cloud with column densities of $N_{\rm H}\sim 10^{23}$\,cm$^{-2}$ impinged by UV radiation from newborn stars appears to best describe the observed line ratios. However, we stress that no single model accurately reproduces simultaneously all of the observed line ratios, thus suggesting that a single cloud is too simplistic a description of the observed phenomenology. }\label{fig_pdr_xdr}
\end{figure*}

\subsection{Physical properties of the cold ISM}\label{sec_ism}

In order to further explore the interplay among CO, \Ci{}, \Cii{} and dust in high-$z$ quasars, we assume that radiative processes (photodissociation, photoelectric heating, etc) dominate over other non--radiative mechanisms altering the energy budget in the molecular gas (shocks, turbulence, cosmic rays, etc). This assumption allows us to directly compare our observations with the line and continuum luminosity predictions of Photo-Dissociation Regions (PDRs) and X-ray Dominated Regions (XDRs). The models are based on the analysis presented in \citet{pensabene21} and briefly summarized here. We use the \textsf{CLOUDY} radiative-transfer code \citep[version c.17.01;][]{ferland17} to predict the line emission of a single plane-parallel semi-infinite cloud impinged by a radiation field in both the PDR and XDR regimes. We created grids of 270 PDR and XDR models with total hydrogen density in the range log($n_{\rm H}$ [cm$^{-3}$]) = [2, 6] (15 values in steps of $\sim$0.29 dex) and strength of the incident radiation field (18 models) in the range $\log G_0$ [Habing units] = [1, 6] (for PDRs), and $\log F_{\rm X}$ [erg\,s$^{-1}$\,cm$^{-2}$] = [-2.0, 2.0] (for XDRs). We consider three cases of total hydrogen column density $\log N_{\rm H}$ [cm$^{-2}$]=\{22, 23, 24\} (see Fig.~\ref{fig_pdr_xdr}). We assume that all the line emission arises from the neutral or molecular gas phase, in particular the contribution of H{\sc ii} regions to the \Cii{} emission is neglected \citep[see, e.g.,][]{diazsantos17,pensabene21}. 

The relative strength of all of the targeted lines is sensitive to the gas density, as well as to the intensity and hardness of the incident radiation field. E.g., the \Ci{}/CO(7--6) ratio is sensitive to the gas density in the regime $n_{\rm H}<10^4$\,cm$^{-3}$, and only marginally affected by other parameters for $N_{\rm H}\gsim 10^{23}$\,cm$^{-2}$. The \Cii{}/\Ci{} typically shows values $>10$ in PDRs, while it is always $<10$ in XDRs, unless $N_{\rm H}\lsim 10^{22}$\,cm$^{-2}$. This is because X-rays penetrate deeper into the clouds than UV photons, thus heating the cloud cores and enhancing \Ci{}. For the same reason, CO molecules receive additional energy and thus the CO spectral line energy distribution shows stronger emission in the high--$J$ transitions. This translates into a suppressed \Cii{}/CO(7--6) luminosity ratio in XDRs compared with predictions for PDRs, at least for high-density clouds ($n_{\rm H}>10^4$\,cm$^{-3}$). At low column densities, $N_{\rm H}\lsim 10^{22}$\,cm$^{-2}$, predictions for this set of observables based on PDR and XDR models tend to converge. The expected \Cii{}/\Ci{} ratios decrease in both models. Most notably, CO(7--6) emission is largely suppressed, thus leading to high \Cii{}/CO(7--6) and \Ci{}/CO(7--6) luminosity ratios.

As discussed in the previous sections, the line luminosities in our sample span a small range ($<$0.5\,dex). This implies that also the probed line ratios show similarly homogeneous values within the sample. We argue that the selection of IR--luminous ($L_{\rm IR}>10^{12}$\,\Lsun{}) sources might bias our sample towards a specific subclass of host galaxies hosting a compact starburst. In Fig.~\ref{fig_pdr_xdr}, we compare the predicted line ratios with the values  observed for the quasars in our sample. We find that PDR models successfully predict the observed \Cii{}/\Ci{} ratios over a wide range of column densities. That is not the case for XDR models, which require low column densities in order to account for the high \Cii{}/\Ci{} luminosity ratios in our data. All models predict a wide range of \Cii{}/CO(7--6) luminosity ratios, although at low column densities it becomes increasingly difficult to explain the low values observed in our study, in particular if compared to predictions from PDR models. Finally, the observed \Ci{}/CO(7--6) luminosity ratios point to intermediate to high densities ($n_{\rm H}\gsim 10^3$\,cm$^{-3}$ in PDRs, $n_{\rm H}\gsim 10^5$\,cm$^{-3}$ in XDRs).

\begin{figure*}
\begin{center}
\includegraphics[width=0.49\textwidth]{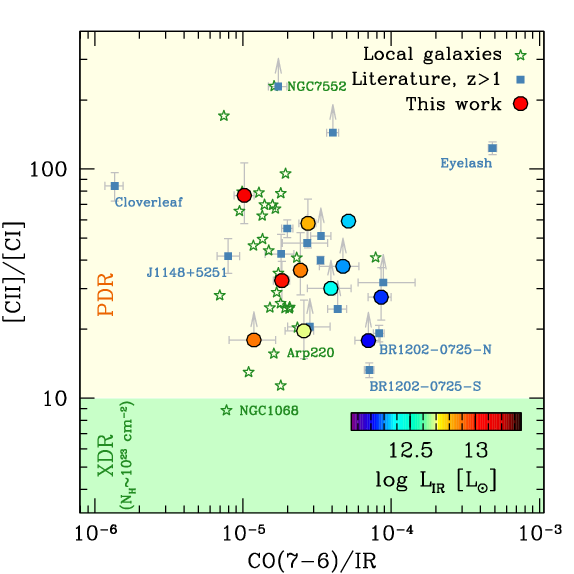}
\includegraphics[width=0.49\textwidth]{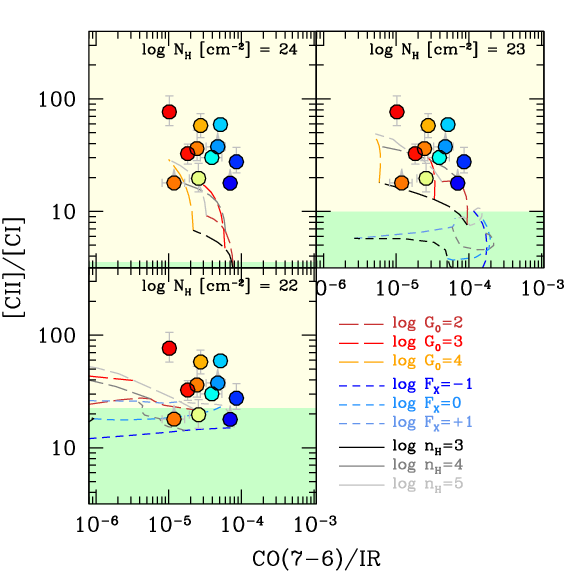}\\
\end{center}
\caption{{\em Left:} The observed \Cii/\Ci{} luminosity line ratio, as a function of the CO(7--6)/IR luminosity ratio, for the quasars in our study (filled circles; see references in the main text), as well as other high redshift ($z>1$) galaxies and quasar hosts (filled squares) and local galaxies (empty stars, based on \citealt{rosenberg15}). All the quasars in our study show \Cii{}/\Ci{} ratios $\gsim20$, which are not compatible with XDRs. Our sample also displays a range of CO(7--6)/IR ratios comparable with other high--$z$ sources, but a factor $\sim 2-3\times$ higher than the bulk of the local galaxies. {\em Right:} Predictions of the \Cii{}/\Ci{} ratio as a function of the CO(7--6)/IR luminosity ratio for various gas densities $\log n_{\rm H} {\rm [cm^{-3}]}=$(3,4,5) for a PDR with incident radiation field $\log G_0=$(2,3,4) and an XDR with $\log F_{\rm X}=$(-1,0,+1). Our sources (shown in filled circled, with the same color scheme as in the left-hand panel) are best described by a PDR--like environment with $N_{\rm H}\sim 10^{23}$\,cm$^{-2}$, $n_{\rm H}\gsim 10^4$\,cm$^{-3}$, and $G_0\sim 10^3$.}\label{fig_ratios}
\end{figure*}

In Fig.~\ref{fig_ratios}, we compare the observed \Cii{}/\Ci{} and CO(7--6)/IR luminosity ratios in individual sources from our sample with other high-redshift quasars and SMGs \citep[from][]{weiss03, riechers09, bradford09, ivison10, cox11, debreuck11, danielson11, wagg12, salome12, walter12, weiss13, riechers13, rawle14, leipski14, decarli14, venemans17a, yang19, zhao20, pensabene21} as well local galaxies \citep[based on the compilation in][]{rosenberg15} and with the predictions from the models presented in Fig.~\ref{fig_pdr_xdr}. The quasars in our study appear to populate a rather narrow parameter space in this diagram: They all have \Cii{}/\Ci{}$\gsim$20, consistent with most of the high-$z$ sources in our comparison, as well as the bulk of the local sample from \citet{rosenberg15}. Notably, only the local Seyfert 2 galaxy NGC\,1068 falls (although, only marginally) in the \Cii{}/\Ci{}$<$10 regime that is often adopted as a {\em bona fide} XDR domain. In terms of CO(7--6)/IR, we find values of $10^{-5}-10^{-4}$, in line with most of the high--$z$ comparison sample, and (on average) a factor 2--3 higher than most of the local galaxies in \citet{rosenberg15}. 

In Fig.~\ref{fig_ratios} we also show how the individual sources presented in our study compare with respect to the predicted \Cii{}/\Ci{} and CO(7--6)/IR luminosity ratios from our PDR and XDR models. As already mentioned, the high values observed for the \Cii{}/\Ci{} ratio point to a relatively low gas column density, $N_{\rm H}\lsim 10^{23}$\,cm$^{-2}$ for PDRs, $\lsim 10^{22}$\,cm$^{-2}$ for XDRs. However, at very low column densities, $N_{\rm H}\sim 10^{22}$\,cm$^{-2}$, the CO(7--6) luminosity is suppressed, thus models fail to reproduce the relatively high CO(7--6)/IR ratios observed in our quasars. The quasar population in our study appears to be best described by PDR models with column densities of $N_{\rm H}\sim 10^{23}$\,cm$^{-2}$, densities of $n_{\rm H}\gsim 10^4$\,cm$^{-3}$, impinged with a radiation field with $G_0\sim 10^3$. These modest column density estimates are in agreement with the estimates based on the dust continuum slope discussed in Sec.~\ref{sec_fits}. At face value, these constraints imply a line-of-sight thickness of the medium of only a few parsecs. However, we stress that the simplified models adopted here (in particular, because of the single-cloud treatment at fixed gas density, and because we neglect non--radiative ingredients contributing to the gas excitation, such as turbulence and cosmic rays) likely fail to capture the complex interplay of different phases in the ISM, thus these findings need to be taken with caution.

\section{Conclusions}\label{sec_conclusions}

We present a survey of \Ci{}, CO(7--6), CO(6--5) and their underlying dust continuum (observed at 3\,mm) in a sample of ten quasar host galaxies at $z\sim 6$. All of the quasars in our sample had previously been detected in their \Cii{} and dust continuum at $\sim 1$\,mm (in the observer's frame). Our main findings are:
\begin{itemize}
\item[{\it i-}] We detect CO(7--6) in all of the targeted quasars, and \Ci{} in 6 of them. Out of the 8 quasar host galaxies for which we have CO(6--5) coverage, we also report six detections in this emission line. The underlying 3\,mm dust continuum is detected in 4/10 of the sample. This work doubles the number of $z\sim 6$ quasars with CO(7--6) measurements in the literature. 
\item[{\it ii-}] The observed flux density of the dust continuum emission at 3\,mm and 1\,mm is consistent with a modified black body emission with temperature $T_{\rm d}= 47$ \,K and an optical depth at the frequency of \Cii{} of $\tau_{\rm 1900\,GHz}=0.2$.
\item[{\it iii-}] Our targets span a factor $\sim 10$ in IR luminosity, but only a factor $\lsim 3$ in the luminosity of any line detected in our study. 
\item[{\it iv-}] We derive molecular gas masses using four independent methods, based on the dust continuum and the CO, \Ci{}, and \Cii{} line emission. To first order, all the methods point to similar values of $M_{\rm H2}=10^{10-11}$\,\Msun{}. This confirms earlier results suggesting that immense gaseous reservoirs power the intense star formation and nuclear activity of the first quasar host galaxies. By combining the four independent measurements, we infer relative constraints on the unknown scaling factors (gas-to-dust, $\alpha_{\rm CO}$, carbon abundance, $\alpha_{\rm [CII]}$, etc). Our estimates for these conversion factors agree with the values inferred in a similar study by \citet{dunne21} on galaxies at $z=0.35$.
\item[{\it v-}] We investigate the physical properties of the cold ISM via \textsf{CLOUDY}-based PDR/XDR models. By comparing the luminosity of dust, \Cii{}, \Ci{}, and CO(7--6), with the expectations from radiative transfer models, we find that the best (albeit imperfect) description of the observations requires a PDR--like environment with dense ($n_{\rm H}>10^4$\,cm$^{-3}$) clouds impinged by a radiation field of $G_0\sim 10^3$ (in Habing units). Column densities of $N_{\rm H}\sim10^{23}$\,cm$^{-2}$ appear favored by our modeling, thus yielding a typical size of the clouds of only a few parsec. 
\end{itemize}
This study confirms and further constrains the presence of massive reservoirs of molecular gas in the host galaxies of $z\sim 6$ quasars, and suggests that star formation is the main mechanism responsible for the gas excitation at galactic scales. This work demonstrates the diagnostic power of combined 3\,mm and 1\,mm observations of very high-redshift sources, that can reveal a wide range of properties of the cold ISM in massive galaxies at cosmic dawn with only two frequency setups. Further NOEMA and ALMA observations will allow us to expand this analysis to a broader range of IR luminosities, and to extend the investigation to other sets of diagnostics (e.g., dense gas tracers, metallicity tracers, etc).

\section*{Acknowledgments}

We are grateful to the referee for their useful and thorough report that helped to improve the manuscript. This work is based on observations carried out under project numbers S052, X04D, S15DA, S17CD, S18DM, S19DM with the IRAM NOEMA Interferometer. IRAM is supported by INSU/CNRS (France), MPG (Germany) and IGN (Spain). The research leading to these results has received funding from the European Union's Horizon 2020 research and innovation program under grant agreement No 730562 [RadioNet]. R.D.\ is grateful to the IRAM staff and local contacts for their help and support in the data processing. F.W.\ and B.P.V.\ acknowledge support from ERC Advanced grant 740246 (Cosmic\_Gas). D.R. acknowledges support from the National Science Foundation under grant numbers AST-1614213 and AST-1910107. D.R. also acknowledges support from the Alexander von Humboldt Foundation through a Humboldt Research Fellowship for Experienced Researchers. F.B.\ acknowledges support through the within the Collaborative Research Centre 956, sub-project A01, funded by the Deutsche Forschungsgemeinschaft (DFG) -- project ID 184018867.

\label{lastpage}


\begin{thebibliography}{99}


\bibitem[Alaghband-Zavena et al.(2013)]{alaghband13} Alaghband-Zadeh S., Chapman S.C., Swinbank A.M., Smail I., Danielson A.L.R., Decarli R., Ivison R.J., Meijerink R., et al.~2013, MNRAS, 435, 1493 
\bibitem[Ba\~{n}ados et al.(2015)]{banados15} Ba\~{n}ados E.,  Decarli R., Walter F., Venemans B.P., Farina E.P., Fan X.~2015, ApJ, 805, L8
\bibitem[Ba\~{n}ados et al.(2018)]{banados18} Ba\~{n}ados E., Venemans B.P., Mazzucchelli C., Farina E.P., Walter F., Wang F., Decarli R., Stern D., et al.~2018, Nature, 553, 473
\bibitem[Beelen et al.(2006)]{beelen06} Beelen A., Cox P., Benford D.J., Dowell C.D., Kov\'{a}cs A., Bertoldi F., Omont A., Carilli C.L.~2006, ApJ, 642, 694
\bibitem[Berta et al.(2016)]{berta16} Berta S., Lutz D., Genzel R., F\"orster-Schreiber N.M., Tacconi L.J., 2016, A\&A, 587, A73
\bibitem[Bertoldi et al.(2003)]{bertoldi03} Bertoldi F., Carilli C.L., Cox P., Fan X., Strauss M.A., Beelen A., Omont A., Zylka R.~2003, A\&A, 406, L55
\bibitem[Beuther et al.(2014)]{beuther14} Beuther H., Ragan S.E., Ossenkopf V., Glover S., Henning Th., Linz H., Nielbock M., Krause O., et al.~2014, A\&A, 571, A53
\bibitem[Bisbas et al.(2019)]{bisbas19} Bisbas T.G., Schruba A., van Dishoeck E.F.~2019, MNRAS, 485, 3097
\bibitem[Bolatto et al.(2013)]{bolatto13} Bolatto A.D., Wolfire M., Leroy A.K., 2013, ARA\&A, 51, 207	
\bibitem[Boogaard et al.(2020)]{boogaard20} Boogaard L.A., van der Werf P.; Weiss A., Popping G., Decarli R., Walter F., Aravena M., Bouwens R., et al.~2020, ApJ, 902, 109
\bibitem[Bothwell et al.(2017)]{bothwell17} Bothwell M.S., Aguirre J.E., Aravena M., Bethermin M., Bisbas T.G., Chapman S.C., De Breuck C., Gonzalez A.H., et al.~2017, MNRAS, 466, 2825
\bibitem[Bradford et al.(2009)]{bradford09} Bradford C.M., Aguirre J.E., Aikin R., Bock J.J., Earle L., Glenn J., Inami H., Maloney P.R., et al.~2009, ApJ, 705, 112
\bibitem[Carilli \& Walter(2013)]{carilli13} Carilli C.L. \& Walter F., 2013, ARA\&A, 51, 105 
\bibitem[Chabrier(2003)]{chabrier03} Chabrier G., 2003, PASP, 115, 763
\bibitem[Clark et al.(2019)]{clark19} Clark P.C., Glover S.C.O., Ragan S.E., Duarte-Cabral A.~2019, MNRAS, 486, 4622
\bibitem[Combes(2018)]{combes18} Combes F.~2018, A\&ARv, 26, 5
\bibitem[Cox et al.(2011)]{cox11} Cox P., Krips M., Neri R., Omont A., G\"usten R., Menten K.M., Wyrowski F., Weiss A., et al., 2011, arXiv:1107.2924
\bibitem[Cubick et al.(2008)]{cubick08} Cubick M.. Stutzki J.. Ossenkopf V., Kramer C., R\"{o}llig M.~2008, A\&A, 488, 623
\bibitem[da Cunha et al.(2013)]{dacunha13} da Cunha E., Groves B., Walter F., Decarli R., Wei\ss{} A., Bertoldi F., Carilli C., Daddi E., Elbaz D., Ivison R., et al., 2013, ApJ, 766, 13
\bibitem[da Cunha et al.(2021)]{dacunha21} da Cunha E., Hodge J.A., Casey C.M., Algera H.S.B., Kaasinen M., Smail I., Walter F., Brandt W.N., et al.~2021, ApJ, 919, 30
\bibitem[Daddi et al.(2010)]{daddi10} Daddi E., Bournaud F., Walter F., Dannerbauer H., Carilli C.L., Dickinson M., Elbaz D., Morrison G.E., et al., 2010a, ApJ, 713, 686
\bibitem[Danielson et al.(2011)]{danielson11} Danielson A.L.R., Swinbank A.M., Smail I., Cox P., Edge A.C., Weiss A., Harris A.I., Baker A.J., et al.~2011, MNRAS, 410, 1687
\bibitem[De Breuck et al.(2011)]{debreuck11} De Breuck C., Maiolino R., Caselli P., Coppin K., Hailey-Dunsheath S., Nagao T.~2011, A\&A, 530, L8 
\bibitem[De Rosa et al.(2014)]{derosa14} De Rosa G., Venemans B.P., Decarli R., Gennaro M., Simcoe R.A., Dietrich M., Peterson B.M., Walter F., et al.~2014, ApJ, 790, 145
\bibitem[Decarli et al.(2014)]{decarli14} Decarli R., Walter F., Carilli C., Riechers D., Cox P., Neri R., Aravena M., Bell E., et al.~2014, ApJ, 782, 78
\bibitem[Decarli et al.(2017)]{decarli17} Decarli R., Walter F., Venemans B.P., Ba\~{n}ados E., Bertoldi F., Carilli C., Fan X., Farina E.P., et al.~2017, Nature, 545, 457
\bibitem[Decarli et al.(2018)]{decarli18} Decarli R., Walter F., Venemans B.P., Ba\~{n}ados E., Bertoldi F., Carilli C., Fan X., Farina E.P., et al.~2018, ApJ, 854, 97
\bibitem[Decarli et al.(2019)]{decarli19} Decarli R., Dotti M., Ba\~{n}ados E., Farina E.P., Walter F., Carilli C., Fan X., Mazzucchelli C., et al.~2019, ApJ, 880, 157
\bibitem[D\'{i}az-Santos et al.(2017)]{diazsantos17} D\'{i}az-Santos T., Armus L., Charmandaris V., Lu N., Stierwalt S., Stacey G., Malhotra S., van der Werf P.P., et al.~2017, ApJ, 846, 32
\bibitem[Drake et al.(2019)]{drake19} Drake A.B., Farina E.P., Neeleman M., Walter F., Venemans B., Ba\~{n}ados E., Mazzucchelli C., Decarli R.~2019, ApJ, 881, 131 
\bibitem[Draine(2003)]{draine03} Draine B.T.~2003, ARA\&A, 41, 241
\bibitem[Drouart et al.(2014)]{drouart14} Drouart G., De Breuck C., Vernet J., Seymour N., Lehnert M., Barthel P., Bauer F.E., Ibar E., et al.~2014, A\&A, 566, A53
\bibitem[Dunne et al.(2021)]{dunne21} Dunne L., Maddox S.J., Vlahakis C., Gomez H.L.~2021, MNRAS, 501, 2573
\bibitem[Farina et al.(2019)]{farina19} Farina E.P., Arrigoni-Battaia F., Costa T., Walter F., Hennawi J.F., Drake A.B., Decarli R., Gutcke T.A., et al.~2019, ApJ, 887, 196
\bibitem[Fabian(2012)]{fabian12} Fabian A., 2012, ARA\&A, 50, 455
\bibitem[Fan et al.(2000)]{fan00} Fan X., White R.L., Davis M., Becker R.H., Strauss M.A., Haiman Z., Schneider D.P., Gregg M.D., et al.~2000, AJ, 120, 1167
\bibitem[Fan et al.(2003)]{fan03} Fan X., Strauss M.A., Schneider D.P., Becker R.H., White R.L., Haiman Z., Gregg M., Pentericci L., et al.~2003, AJ, 125, 1649
\bibitem[Ferland et al.(2017)]{ferland17} Ferland G.J., Chatzikos M., Guzm\'{a}n F., Lykins M.L., van Hoof P.A.M., Williams R.J.R., Abel N.P., Badnell N.R., et al.~2017, RMxAA, 53, 385
\bibitem[Frerking et al.(1989)]{frerking89} Frerking M.A., Keene J., Blake G.A., Phillips T.G.~1989, ApJ, 344, 311
\bibitem[Genzel et al.(2015)]{genzel15} Genzel R., Tacconi L.J., Lutz D., Saintonge A., Berta S., Magnelli B., Combes F., Garc\'{i}a-Burillo S., Neri R., et al.~2015, ApJ, 800, 20
\bibitem[Goicoechea et al.(2019)]{goicoechea19} Goicoechea J.R., Santa-Maria M. G., Bron E., Teyssier D., Marcelino N., Cernicharo J., Cuadrado S.~2019, A\&A, 622, A91
\bibitem[Glover \& Clark(2016)]{glover16} Glover S.C.O.~\& Clark P.C.~2016, MNRAS, 456, 3596
\bibitem[Greve et al.(2014)]{greve14} Greve T.R., Leonidaki I., Xilouris E.M., Wei\ss{} A., Zhang Z.-Y., van der Werf P., Aalto S., Armus L., et al.~2014, ApJ, 794, 142
\bibitem[Groves et al.(2015)]{groves15} Groves B.A., Schinnerer E., Leroy A., Galametz M., Walter F., Bolatto A., Hunt L., Dale D., et al.~2015, ApJ, 799, 96
\bibitem[Gullberg et al.(2015)]{gullberg15} Gullberg B., De Breuck C., Vieira J.D., Wei\ss{} A., Aguirre J.E., Aravena M., B\'{e}thermin M., Bradford C.M., et al.~2015, MNRAS, 449, 2883
\bibitem[Heckman \& Best(2014)]{heckman14} Heckman T.M., Best P.N.~2014, ARA\&A, 52, 589
\bibitem[Herrera-Camus et al.(2018)]{herreracamus18} Herrera-Camus R., Sturm E., Graci\'{a}-Carpio J., Lutz D., Contursi A., Veilleux S., Fischer J., Gonz\'{a}lez-Alfonso E., Poglitsch A., et al.~2018, ApJ, 861, 95
\bibitem[Hodge \& da Cunha (2020)]{hodge20} Hodge J.A.\ \& da Cunha E.~2020 (arXiv:2004.00934)
\bibitem[Ivison et al.(2010)]{ivison10} Ivison R.J., Swinbank A.M., Swinyard B., Smail I., Pearson C.P., Rigopoulou D., Polehampton E., Baluteau J.-P., et al., 2010, A\&A, 518, L35
\bibitem[Kennicutt \& Evans(2012)]{kennicutt12} Kennicutt R.C. \& Evans N.J., 2012, ARA\&A, 50, 531
\bibitem[King \& Pounds(2015)]{king15} King A., Pounds K.~2015, ARA\&A, 53, 115
\bibitem[Kramer et al.(2004)]{kramer04} Kramer C., Jakob H., Mookerjea B., Schneider N., Br\"{u}ll M., Stutzki J.~2004, A\&A, 424, 887
\bibitem[Leipski et al.(2014)]{leipski14} Leipski C., Meisenheimer K., Walter F., Klaas U., Dannerbauer H., De Rosa G., Fan X., Haas M., et al.~2014, ApJ, 785, 154
\bibitem[Li et al.(2020)]{li20} Li J., Wang R., Riechers D., Walter F., Decarli R., Venemans B.P., Neri R., Shao Y., et al.~2020, ApJ, 889, 162
\bibitem[Maloney et al.(1996)]{maloney96} Maloney P.R., Hollenbach D.J., Tielens A.G.G.M.~1996, ApJ, 466, 561
\bibitem[Mazzucchelli et al.(2017)]{mazzucchelli17} Mazzucchelli C., Ba\~{n}ados E., Venemans B.P., Decarli R., Farina E.P., Walter F., Eilers A.-C., Rix H.-W., et al.~2017, ApJ, 849, 91
\bibitem[Meijerink et al.(2007)]{meijerink07} Meijerink R., Spaans M., Israel F.P.~2007, A\&A, 461, 793
\bibitem[Meyer et al.(2022)]{meyer22} Meyer R.A., et al., 2022, ApJ, submitted
\bibitem[Neeleman et al.(2021)]{neeleman21} Neeleman M., Novak M., Venemans B.P., Walter F., Decarli R., Kaasinen M., Schindler J.-T., Ba\~{n}ados E., et al.~2021, ApJ, 911, 141
\bibitem[Novak et al.(2019)]{novak19} Novak M., Ba\~{n}ados E., Decarli R., Walter F., Venemans B., Neeleman M., Farina E.P., Mazzucchelli C., et al.~2019, ApJ, 881, 63  
\bibitem[Papadopoulos et al.(2018)]{papadopoulos18} Papadopoulos P.P., Bisbas T.G., Zhang Z.-Y.~2018, MNRAS, 478, 1716
\bibitem[Pensabene et al.(2021)]{pensabene21} Pensabene A., Decarli R., Ba\~{n}ados E., Venemans B., Walter F., Bertoldi F., Fan X., Farina E.P., et al.~2021, A\&A, 652, A66
\bibitem[Popping et al.(2017)]{popping17} Popping G., Decarli R., Man A.W.S., Nelson E.J., B\'{e}thermin M., De Breuck C., Mainieri V., van Dokkum P.G., et al.~2017, A\&A, 602, A11
\bibitem[Rawle et al.(2014)]{rawle14} Rawle T.D., Egami E., Bussmann R.S., Gurwell M., Ivison R.J., Boone F., Combes F., Danielson A.L.R., et al.~2014, ApJ, 783, 59
\bibitem[Riechers et al.(2009)]{riechers09} Riechers D.A., Walter F., Bertoldi F., Carilli C.L., Aravena M., Neri R., Cox P., Wei\ss{} A., et al.~2009, ApJ, 703, 1338
\bibitem[Riechers et al.(2013)]{riechers13} Riechers D.A., Bradford C.M., Clements D.L., Dowell C.D., P\'{e}rez-Fournon I., Ivison R.J., Bridge C., Conley A., et al. 2013 Nature, 496, 329
\bibitem[Rosenberg et al.(2015)]{rosenberg15} Rosenberg M.J.F., van der Werf P.P., Aalto S., Armus L., Charmandaris V., D\'{i}az-Santos T., Evans A.S., Fischer J., et al.~2015, ApJ, 801, 72
\bibitem[Salom\'{e} et al.(2012)]{salome12} Salom\'{e} P., Gu\'{e}lin M., Downes D., Cox P., Guilloteau S., Omont A., Gavazzi R., Neri R.~2012, A\&A, 545, A57   
\bibitem[Salpeter(1955)]{salpeter55} Salpeter E.E., 1955, ApJ, 121, 161
\bibitem[Sandstrom et al.(2013)]{sandstrom13} Sandstrom K.M., Leroy A.K., Walter F., Bolatto A.D., Croxall K.V., Draine B.T., Wilson C.D., Wolfire M., et al.~2013, ApJ, 777, 5
\bibitem[Schindler et al.(2020)]{schindler20} Schindler J.-T., Farina E.P., Ba\~{n}ados E., Eilers A.-C., Hennawi J.F., Onoue M., Venemans B.P., Walter F., et al.~2020, ApJ, 905, 51
\bibitem[Scoville et al.(2017a)]{scoville17} Scoville N., Lee N., Vanden Bout P., Diaz-Santos T., Sanders D., Darvish B., Bongiorno A., Casey C.M., et al.~2017a, ApJ, 837, 150
\bibitem[Scoville et al.(2017b)]{scoville17b} Scoville N., Murchikova L., Walter F., Vlahakis C., Koda J., Vanden Bout P., Barnes J., Hernquist L., et al.~2017b, ApJ, 836, 665
\bibitem[Somerville \& Dav\'{e}(2015)]{somerville15} Somerville R.S., Dav\'{e} R.~2015, ARA\&A, 53, 51
\bibitem[Sommovigo et al.(2021)]{sommovigo21} Sommovigo L., Ferrara A., Carniani S., Pallottini A., Gallerani S., Vallini L.~2021, MNRAS, 503, 4878
\bibitem[Sommovigo et al.(2022)]{sommovigo22} Sommovigo L., Ferrara A., Pallottini A.,  Dayal P., Bouwens R.J., Smit R., da Cunha E., De Looze, I., et al.~2022, MNRAS, in press
\bibitem[Stacey et al.(2010)]{stacey10} Stacey G.J., Hailey-Dunsheath S., Ferkinhoff C., Nikola T., Parshley S.C., Benford D.J., Staguhn J.G., Fiolet N., 2010, ApJ, 724, 957
\bibitem[Sutter et al.(2021)]{sutter21} Sutter J., Dale D.A., Sandstrom K., Smith J.D.T., Bolatto A., Boquien M., Calzetti D., Croxall K.V., et al.~2021, MNRAS, 503, 911
\bibitem[Tacconi et al.(2020)]{tacconi20} Tacconi L.J., Genzel R., Sternberg A.~2020 (arXiv:2003.06245)
\bibitem[Tielens \& Hollenbach(1985)]{tielens85} Tielens A.G.G.M., Hollenbach D.~1985, ApJ, 291, 722
\bibitem[Trakhtenbrot et al.(2017)]{trakhtenbrot17} Trakhtenbrot B., Lira P., Netzer H., Cicone C., Maiolino R., Shemmer O.~2017, ApJ, 836, 8
\bibitem[Valentino et al.(2018)]{valentino18} Valentino F., Magdis G.E., Daddi E., Liu D., Aravena M., Bournaud F., Cibinel A., Cormier D., et al.~2018, ApJ, 869, 27
\bibitem[Valentino et al.(2020)]{valentino20} Valentino F., Magdis G.E., Daddi E., Liu D., Aravena M., Bournaud F., Cortzen I., Gao Y., et al.~2020, ApJ, 890, 24
\bibitem[Vallini et al.(2019)]{vallini19} Vallini L., Tielens A.G.G.M., Pallottini A., Gallerani S., Gruppioni C., Carniani S., Pozzi F., Talia M.~2019, MNRAS, 490, 4502
\bibitem[Venemans et al.(2017a)]{venemans17a} Venemans B.P., Walter F., Decarli R., Ferkinhoff C., Wei\ss{} A., Findlay J.R., McMahon R.G., Sutherland W.J., Meijerink R., 2017a, ApJ, 845, 154
\bibitem[Venemans et al.(2017b)]{venemans17b} Venemans B.P., Walter F., Decarli R., Ba\~{n}ados E., Carilli C., Winters J.M., Schuster K., da Cunha E., et al.~2017b, ApJ, 851, L8
\bibitem[Venemans et al.(2018)]{venemans18} Venemans B.P., Decarli R., Walter F., Ba\~{n}ados E., Bertoldi F., Fan X., Farina E.P., Mazzucchelli C., et al.~2018, ApJ, 866, 159
\bibitem[Venemans et al.(2020)]{venemans20} Venemans B.P., Walter F., Neeleman M., Novak M., Otter J., Decarli R., Ba\~{n}ados E., Drake A., et al.~2020, ApJ, 904, 130 
\bibitem[Vito et al.(2019)]{vito19} Vito F., Brandt W.N., Bauer F.E., Gilli R., Luo B., Zamorani G., Calura F., Comastri A., et al.~2019, A\&A, 628, L6
\bibitem[Wagg et al.(2012)]{wagg12} Wagg J., Wiklind T., Carilli C.L., Espada D., Peck A., Riechers D., Walter F., Wootten A., et al.~2012, ApJ, 752, L30
\bibitem[Walter et al.(2003)]{walter03} Walter F., Bertoldi F., Carilli C., Cox P., Lo K.Y., Neri R., Fan X., Omont A., Strauss M.A., Menten K.M.~2003, Nature, 424, 406
\bibitem[Walter et al.(2011)]{walter11} Walter F., Weiss A., Downes D., Decarli R., Henkel C.~2011, ApJ, 730, 18
\bibitem[Walter et al.(2012)]{walter12} Walter F., Decarli R., Carilli C., Bertoldi F., Cox P., da Cunha E., Daddi E., Dickinson M., et al., 2012, Nature, 486, 233
\bibitem[Walter et al.(2022)]{walter22} Walter F., Neeleman M., Decarli R., Venemans B., Meyer R., Weiss A., Ba\~{n}ados E., Bosman S.E.I., et al.~2022, ApJ, in press (arXiv:2201.06396)
\bibitem[Wang et al.(2008a)]{wang08a} Wang R., Wagg J., Carilli C.L., Benford D.J., Dowell C.D., Bertoldi F., Walter F., Menten K.M., et al.~2008, AJ, 135, 1201
\bibitem[Wang et al.(2008b)]{wang08b} Wang R., Carilli C.L., Wagg J., Bertoldi F., Walter F., Menten K.M., Omont A., Cox P., et al.~2008b, ApJ, 687, 848
\bibitem[Wang et al.(2010)]{wang10} Wang R., Carilli C.L., Neri R., Riechers D.A., Wagg J., Walter F., Bertoldi F., Menten K.M., et al.~2010, ApJ, 714, 699
\bibitem[Wang et al.(2011)]{wang11} Wang R., Wagg, J., Carilli C.L., Neri R., Walter F., Omont A., Riechers D.A., Bertoldi F., et al.~2011, AJ, 142, 101
\bibitem[Wang et al.(2013)]{wang13} Wang R., Wagg J., Carilli C.L., Walter F., Lentati L., Fan X., Riechers D.A., Bertoldi F., Narayanan D., Strauss M.A. et al.~2013, ApJ, 773, 44
\bibitem[Wang et al.(2019)]{wang19} Wang F., Wang R., Fan X., Wu X.-B., Yang J., Neri R., Yue M.~2019, ApJ, 880, 2
\bibitem[Wang et al.(2021)]{wang21} Wang F., Yang J., Fan X., Hennawi J.F., Barth A.J., Banados E., Bian F., Boutsia K., et al.~2021, ApJ, 907, L1
\bibitem[Wei\ss{} et al.(2003)]{weiss03} Wei\ss{} A., Henkel C., Downes D., Walter F.~2003, A\&A, 409, L41
\bibitem[Wei\ss{} et al.(2005)]{weiss05} Wei\ss{} A., Downes D., Henkel C., Walter F.~2005, A\&A, 429, L25
\bibitem[Wei\ss{} et al.(2007)]{weiss07} Wei\ss{} A., Downes D., Neri R., Walter F., Henkel C., Wilner D.J., Wagg J., Wiklind T., 2007, A\&A, 467, 955
\bibitem[Wei\ss{} et al.(2013)]{weiss13} Wei\ss{} A., De Breuck C., Marrone D.P., Vieira J.D., Aguirre J.E., Aird K.A., Aravena M., Ashby M.L.N., et al., 2013, ApJ, 767, 88
\bibitem[Willott et al.(2017)]{willott17} Willott C.J., Bergeron J., Omont A.~2017, ApJ, 850, 108
\bibitem[Yang et al.(2019)]{yang19} Yang J., Venemans B., Wang F., Fan X., Novak M., Decarli R., Walter F., Yue M., et al.~2019, ApJ, 880, 153
\bibitem[Yang et al.(2021)]{yang21} Yang J., Wang F., Fan X., Barth A.J., Hennawi J.F., Nanni R., Bian F., Davies F.B., et al.~2021, ApJ, in press (arXiv:2109.13942)
\bibitem[Yang et al.(2020)]{yang20} Yang J., Wang F., Fan X., Hennawi J.F., Davies F.B., Yue Mi., Banados E., Wu X.-B., et al.~2020, ApJ, 897, L14
\bibitem[Zanella et al.(2018)]{zanella18} Zanella A., Daddi E., Magdis G., Diaz Santos T., Cormier D., Liu D., Cibinel A., Gobat R., et al.~2018, MNRAS, 481, 1976
\bibitem[Zhao et al.(2020)]{zhao20} Zhao Y., Lu N., D\'{i}az-Santos T., Gao Y., Xu C.K., Charmandaris V., Inami H., Rigopoulou D., et al.~2020, ApJ, 892, 145


%
\end{thebibliography}
\end{document}